\documentclass[12pt,a4paper]{iopart}
\usepackage{iopams}
\usepackage{epsfig}
\usepackage{url}
\newcommand{\eqref}[1]{(\ref{#1})}

\newcommand{\zrec}{z_{\rm rec}}

\begin{document}

\title{Confronting Lemaitre-Tolman-Bondi models with Observational Cosmology}

\author{Juan Garcia-Bellido$^{1,2}$, Troels Haugb{\o}lle$^{1,3}$}
\address{$^1$ Instituto de F\'{\i}sica Te\'{o}rica UAM-CSIC,
Universidad Aut\'{o}noma de Madrid, Cantoblanco, 28049 Madrid, Spain,\\
$^2$ Kavli Institute for Theoretical Physics, University of California
Santa Barbara, CA 93106-4030,\\
$^3$ Department of Physics and Astronomy, University of Aarhus, DK-8000
Aarhus C, Denmark} 
\ead{juan.garciabellido@uam.es, haugboel@phys.au.dk}

\begin{abstract}
The possibility that we live in a special place in the universe, close
to the centre of a large void, seems an appealing alternative to the
prevailing interpretation of the acceleration of the universe in terms
of a $\Lambda$CDM model with a dominant dark energy component. In this
paper we confront the asymptotically flat Lema\^itre-Tolman-Bondi (LTB)
models with a series of observations, from Type Ia Supernovae to
Cosmic Microwave Background and Baryon Acoustic Oscillations data.  We
propose two concrete LTB models describing a local void in which the
only arbitrary functions are the radial dependence of the matter density
$\Omega_M$ and the Hubble expansion rate $H$. We find that all observations
can be accommodated within 1
sigma, for our models with 4 or 5 independent parameters.  The best
fit models have a $\chi^2$ very close to that of the $\Lambda$CDM
model. A general Fortran program for comparing LTB models
with cosmological observations, that has been used to make the
parameter scan in this paper, is made public, and can be downloaded at
\url{http://www.phys.au.dk/~haugboel/software.shtml} together with
{\tt IDL} routines for creating the Likelihood plots. 
We perform a simple Bayesian analysis and show that one cannot
exclude the hypothesis that we live within a large local void of an
otherwise Einstein-de Sitter model.\\
\end{abstract}
\pacs{98.80.Cq \hspace{\stretch{1}} Preprint: IFT-UAM/CSIC-08-03,NSF-KITP-08-02}
\submitto{JCAP}
\maketitle

\section{Introduction}

Cosmology has been traditionally prone to speculation, a science which
has had more of wishful thinking than actual deduction. Even today,
with the extraordinary revolution in observational cosmology, there
are many assumptions we have to take for granted, in order to {\em
interpret} the present observations. In particular, the present
abundance of less that 5\% of critical density in matter we know and
measure in the laboratory, i.e.~the baryons we are made of, compared
to the 23\% of unknown ``dark matter'' and the 72\% of even more
unknown ``dark energy'', seems a rather peculiar and perhaps
suspicious composition for the universe, given their disparate evolution
rates. Are we sure we have now the correct picture of the universe?

We live in a very isolated part of the universe and have not reached
that much beyond our solar system. All we can say about the universe
has been inferred from observations done under very general
assumptions; but some of them, no matter how fundamental they may
seem, may be wrong. It is worth pointing out how Einstein failed to
make one of the most fundamental predictions of his new theory
of general relativity, and had to introduce an ``absolute'' in the
theory $-$ the cosmological constant $-$ much to his regret, because he
had the prejudice, supported by incomplete observational data, that we
lived in a static universe without a beginning or an end. Soon after
the discovery of the ``recession'' of galaxies by Hubble did he
renounce to his prejudice on the validity of the Perfect Cosmological
Principle, which assumed maximally symmetric space-times. For 70 years
cosmologists have worked under the assumption of the less strict
Cosmological Principle, which imposes maximal symmetry (homogeneity
and isotropy) only for the spatial sections. But are these symmetries
consistent with observations? It is evident to anyone that looks at
the sky in a clear night far away from city lights that the universe
is {\em not} homogeneous and isotropic.

It has usually been argued that these fundamental symmetries should
only be ``expected to be valid on very large scales'', but how {\em
large} are large scales?  In fact, the distribution of matter in our
local vicinity, i.e.~within several Mpc, is very far from homogeneous;
density contrasts reach enormous values not only at the centres of
galaxies but also on larger scales like clusters and superclusters,
stretching over hundreds of Mpc. So, how far do we have to go before
we reach homogeneity? Certainly present galaxy catalogues are not really
there yet, and it has been speculated that perhaps with the next
generation of deep catalogues like DES \cite{des} we will finally reach the
homogeneity limit. But, if we do {\em not} live in a homogeneous
universe, how do we interpret the observations we have of objects
whose light has travelled a significant fraction of the age of the
universe in order to reach us? Our present assumption is that, in
practice, the intervening inhomogeneity averages out and everything
works as if we lived in a homogeneous universe. For several decades
this assumption has been a valid one, and has provided confidence into
the construction of the so-called Standard Model of Cosmology.

It is only now that we begin to have sufficiently good cosmological
data, and certainly will have even better data in the near future,
that we can be critical and pose the appropriate questions. The issue
of homogeneity of the universe has been often dismissed because of the
apparent extraordinary isotropy of the cosmic microwave background.
However, any mathematician can readily show that isotropy and
homogeneity are very different sets of symmetries and one does not
imply the other. Nevertheless, if we impose the further 
assumption, usually stated as the Copernican Principle, that any
point in space should be equivalent to any other, i.e.~that we don't
live in any special place in the universe, then a mathematical theorem
states that if {\em all} equal observers see the universe isotropic around
them, then the universe must be not only isotropic but also homogeneous.
What remains to be proven is that all observers are in fact equivalent
in the patch of the universe we call the observable universe. Some
will be near a large concentration of mass and others will be in large
voids. Certainly what these two types of observers see will differ
from what an idealised observer living in a perfectly homogeneous
universe would see. Unfortunately, we have never spoken to anyone at
the other side of the universe.

The advantage of the present state of Cosmology is that we can begin
to pose those questions and hope to get concrete answers, while just
ten years ago it would have been futile. Moreover, like with the first
inclusion of the cosmological constant in the theory, almost 100 years
ago, the physics community is very much puzzled about the nature of
this so-called vacuum energy. Its properties defy our basic
understanding of quantum physics and, moreover, it reminds us
suspiciously of the Maxwellian ether, which led the way (via its
disappearance) to a new understanding of physical reality. To put the
question straight, are we sure we live in an accelerated universe
which is driven by some unknown vacuum energy? Could it be that we
have {\em misinterpreted} our superb cosmological data and what those
photons coming from afar are telling us is something completely
different? 

The last few years we have seen a tremendous burst of activity, both at
the theoretical and observational level, in order to disentangle the
subtle intricacies of the actual data sets from their interpretation. As
with any hard science (and Cosmology is indeed finally becoming one),
such enterprises can be approached only via further observational
crosschecks. It is no longer true that ``astrophysicists are often in
error, but never in doubt'', as Lev Landau once said to annoy his
colleague Yakov Zel'dovich \cite{landau}.
We can now propose new ways to measure more observables in a wider
theoretical construction. Perhaps it is time to explore the troubled
waters of non-maximally symmetric spatial sections of the universe.
In particular, since we indeed observe a high degree of isotropy in 
the cosmic microwave background, we can start by exploring the
simplified version of a spherically symmetric inhomogeneous model,
which comes under the name of the Lema\^itre-Tolman-Bondi
(LTB) model~\cite{Lemaitre:1997ab,Tolman:1934za,Bondi:1947av}.

While ordinary Friedmann-Robertson-Walker (FRW) space-times are
characterised by two functions, the Hubble rate $H(t)$ and the density
parameter $\Omega(t)$, which depend on cosmic time but are independent
of the radial coordinate, the LTB models have also two arbitrary
functions, $H(r,t)$ and $\Omega(r,t)$, which depend on both time and
the radial coordinate. The Einstein equations for the LTB model are
sufficiently simple that they can be integrated exactly in terms of
two arbitrary boundary conditions. Thus, the LTB models have lower
symmetries, and thus more freedom, but nevertheless make specific
predictions about the behaviour of light along geodesics of the new
metric. Therefore, one can evaluate the corresponding observables,
like cosmological distances (or ratios of distances), from the
available sets of data. Surprisingly, as we will show in the next
sections, present cosmological data does not yet seem able to exclude
with confidence a universe which is not exactly homogeneous. This
should not come as a surprise, we {\em do} live in an (locally)
inhomogeneous universe, some say that within a large underdense 
void~\cite{Zehavi:1998gz,Tomita:2000rf,Tomita:2000jj,Tomita:2001gh,Frith:2003tb,Busswell:2003ta} similar to 
that which induces a cold spot in the 
CMB~\cite{Vielva:2003et,Cruz:2006sv,Cruz:2006fy}
so why should we assume global homogeneity? In fact, the possibility 
that we happen to live in the centre of the world was advocated
long ago in Ref.~\cite{Linde:1994gy}, based on the stochastic 
inflation formalism~\cite{Linde:1993xx,GarciaBellido:1993wn}.
What is surprising is that
within this framework one can account for (almost) all observational
evidence without having to introduce an unknown in the theory, an
absolute, whose properties are highly mysterious, not to mention the
alternative highly artificial and ad hoc modifications of gravity on
the very large scales.

In this paper we propose an inhomogeneous model of the universe, 
with a local void size of a few Gpc and asymptotically Einstein-de 
Sitter, and then use the present sets of data (CMB, LSS, BAO, SNIa,
HST, Ages, etc.) to constrain its parameters. In Section~2 we describe 
the general LTB metric, the Einstein equations and the definitions of
cosmic distances.  We also give a novel series expansion that allows
one to integrate to arbitrary precision the Einstein equations for arbitrary
functions $H_0(r)$ and $\Omega_M(r)$. We then describe a specific
model with a concrete form of these two functions that are plausible but
simple matches to observations, and show the different (longitudinal and
transverse) rates of expansion and the apparent acceleration. We also
show that the comoving size of the sound horizon depends on the distance
to the centre of the void in the LTB model, and relate it to the
expansion rates at different redshifts. This is a prerequisite for making
a realistic comparison with Baryon Acoustic Oscillation data. In
Sect.~3 we present the data sets we use to constrain the model,
with the subtleties needed to correctly interpret these in terms of
the model. In Sect.~4 we give the main results and present a Bayesian
analysis to ascertain the goodness of fit of the model by comparing it
with a standard $\Lambda$CDM model with variable (but constant)
equation of state parameter $w$. In Sect.~5 we give the conclusions.

\section{The LTB model}

The most general (cosmological) metric satisfying spherically
symmetric spatial sections can be written as
\begin{equation}
ds^2 = - dt^2 + X^2(r,t)\,dr^2 + A^2(r,t)\,d\Omega^2\,,
\end{equation}
where $d\Omega^2 = d\theta^2 + \sin^2\theta d\phi^2$.
Assuming a spherically symmetric matter source,
$$T^\mu_\nu = - \rho_M(r,t)\,\delta^\mu_0\,\delta^0_\nu\,,$$
the $(0,r)$ component of the Einstein equations, $G^0_r = 0$, implies
$X(r,t)=A'(r,t)/\sqrt{1-k(r)}$, with an arbitrary function $k(r)$
playing the role of the spatial curvature parameter. Note that we
recover the FRW metric imposing the extra homogeneity conditions,
$$A(r,t) = a(t)\,r\,, \hspace{2cm} k(r) = k\,r^2\,.$$
The other components of Einstein equations read\cite{Enqvist:2006cg,Enqvist:2007vb}
\begin{eqnarray}
&&{\dot A^2 + k\over A^2} + 2{\dot A\dot A'\over AA'} + {k'(r)\over
A A'} = 8\pi\,G\,\rho_M \,, \\
&&\dot A^2 + 2A\ddot A + k(r) = 0\,.
\end{eqnarray}
Integrating the last equation, we get
\begin{equation}
{\dot A^2\over A^2} = {F(r)\over A^3} - {k(r)\over A^2}\,,
\end{equation}
with another arbitrary function $F(r)$, playing the role of effective
matter content, which substituted into the first equation gives
\begin{equation}
{F'(r)\over A'A^2(r,t)} = 8\pi\,G\,\rho_M(r,t)\,.
\end{equation}
Combining these two equations we arrive at
\begin{equation}
{2\over3}{\ddot A\over A} + {1\over3}{\ddot A'\over A'} =
- {4\pi\,G\over3}\,\rho_M\,,
\end{equation}
which determines the ``effective" acceleration in these inhomogeneous
cosmologies. Note that the notion of acceleration becomes ambiguous
since backward proper time has both a time and a spatial component.

The boundary condition functions $F(r)$ and $k(r)$ are specified by
the nature of the inhomogeneities through the local Hubble rate, the
local matter density and the local spatial curvature,
\begin{eqnarray}
&&H(r,t) = {\dot A(r,t) \over A(r,t)} \,, \\
&&F(r) = H_0^2(r)\,\Omega_M(r)\,A_0^3(r)\,, \\[2mm]
&&k(r) = H_0^2(r)\Big(\Omega_M(r)-1\Big)\,A_0^2(r) \,,
\end{eqnarray}
where functions with subscripts $0$ correspond to present values,
$A_0(r) = A(r,t_0)$ and $H_0(r) = H(r,t_0)$. With these definitions, the
$r$-dependent Hubble rate is written as\cite{Enqvist:2006cg,Enqvist:2007vb}
\begin{equation}\label{eq:hubblerate}
H^2(r,t) = H_0^2(r)\left[\Omega_M(r)\left({A_0(r)\over A(r,t)}\right)^3 +
(1-\Omega_M(r))\left({A_0(r)\over A(r,t)}\right)^2\right]\,.
\end{equation}

Inhomogeneities come in two different classes: in the matter distribution
or in the expansion rate, which are mutually independent. Moreover,
the extra gauge
freedom of the synchronous comoving gauge allows us to choose
\begin{equation}
A(r,t_0) = A_0(r) = r\,.
\end{equation}
Then we can integrate the Hamiltonian constraint equation
(\ref{eq:hubblerate}) to provide comoving time as a function of the
radial coordinate,
\begin{eqnarray}
&&
\hspace*{-2cm}
H_0(r)t(r) = \int^{A(r,t)/A_0(r)} {dx\over\sqrt{\Omega_M(r)/x + 1 -
\Omega_M(r)}} \nonumber \\
&&
\hspace*{-1cm}
= {A(r,t)\over A_0(r)\sqrt{\Omega_K(r)}}\sqrt{1 + {\Omega_M(r)A_0(r)\over
\Omega_K(r)A(r,t)}} - {\Omega_M(r)\over\sqrt{\Omega_K^3(r)}}\ {\rm sinh}^{-1}
\sqrt{\Omega_K(r)A(r,t)\over\Omega_M(r)A_0(r)}\,,\label{TBB}
\end{eqnarray}
where $\Omega_K(r)=1-\Omega_M(r)$. In particular, setting $A(r,t)=A_0(r)$
we find the current age of the universe
\begin{equation}
H_0(r)t_{\rm BB}(r) = {1\over \sqrt{\Omega_K(r)}}\sqrt{1 + {\Omega_M(r)\over
\Omega_K(r)}} - {\Omega_M(r)\over\sqrt{\Omega_K^3(r)}}\ {\rm sinh}^{-1}
\sqrt{\Omega_K(r)\over\Omega_M(r)}\,.\label{TBB2}
\end{equation}

For an observer located at the centre $r=0$, by symmetry, incoming light travels
along radial null geodesics, $ds^2=d\Omega^2=0$, and time decreases
when going away, $dt/dr < 0$, and we have
\begin{equation}
{dt\over dr} = - {A'(r,t)\over\sqrt{1-k(r)}}
\end{equation}
which, together with the redshift equation,
\begin{equation}\label{eq:null}
{d\log(1+z)\over dr} = {\dot A'(r,t)\over\sqrt{1-k(r)}}
\end{equation}
can be written as a parametric set of differential equations,
with $N=\log(1+z)$ being the effective number of e-folds
before the present time, 
\begin{eqnarray}
&&{dt\over dN} = - {A'(r,t)\over\dot A'(r,t)} \,,\\
&&{dr\over dN} = {\sqrt{1-k(r)}\over\dot A'(r,t)} \,,
\end{eqnarray}
from which the functions $t(z)$ and $r(z)$ can be obtained.
From there one can immediately obtain both the luminosity distance,
the comoving distance and the angular diameter distance as a 
function of redshift,
\begin{eqnarray}
&&d_L(z) = (1+z)^2A[r(z),t(z)]\,, \\
&&d_C(z) = (1+z)\,A[r(z),t(z)]\,, \\
&&d_A(z) = A[r(z),t(z)]\,,
\end{eqnarray}

\subsection{Series solution}

In order to integrate out the redshift dependence it will be useful
to make a series expansion of the cosmic time variable $t(r)$ as a
function of the space-dependent scale factor $A(r,t)$. For this
purpose we will define new variables
\begin{eqnarray}
&&y = {\Omega_K(r)\over\Omega_M(r)}\,
H_0(r) \sqrt{\Omega_K(r)}\ t(r) = {2\over3}(\delta\,a)^{3/2}\,, \\
&&x = {\Omega_K(r)\over\Omega_M(r)}\,{A(r,t)\over A_0(r)} = \delta\,{A\over A_0}\,,
\end{eqnarray}
where $\delta(r)$ is a generalised density contrast ratio and $a(t)$ 
is the Einstein-de Sitter (FRW) scale factor,
\begin{eqnarray}
&&\delta(r) = {\Omega_K(r)\over\Omega_M(r)}\,, \\
&&\ a(t) = \left({3\over2}\,H_0(r)\,\sqrt{\Omega_M(r)}\ t\right)^{2/3}\,.
\end{eqnarray}
With these definitions, the time integral (\ref{TBB}) can be written as
$y = \sqrt{x(1+x)} - \ln[\sqrt{x} + \sqrt{1+x}]$, which can be expanded in
series and inverted. With the definitions
\begin{eqnarray}
&&g(z) = z + {1\over5} z^2 - {3\over175} z^3 + {23\over7875} z^4 -
{1894\over3031875} z^5 + {\cal O}(z^6)\,, \\[2mm]
&&f(r) = {H_0'(r)\over H_0(r)} - {\Omega_M'(r)\over\Omega_M(r)}
\left({1+\Omega_M(r)/2\over1-\Omega_M(r)}\right)\,,\\[1mm]
&&h(r) = {1\over r} + {\Omega_M'(r)\over\Omega_M(r)(1-\Omega_M(r))}\,,
\end{eqnarray}
we can write the solution as a power series, whose coefficients can be
calculated with arbitrary precision,
\begin{eqnarray}
&&\hspace*{-1.5cm}
A(r,t) = {r\over \delta}\,g(a\,\delta)\,,\\
&&\hspace*{-1.5cm}
\dot A(r,t) = {2\over3t}{r\over \delta}\,a\,\delta\,g'(a\,\delta)\,,\\
&&\hspace*{-1.5cm}
A'(r,t) = {r\over \delta}\left[g(a\,\delta)h(r) +
{2\over3}\,a\,\delta\,g'(a\,\delta)\,f(r)\right]\,,\\
&&\hspace*{-1.5cm}
\dot A'(r,t) = {2\over3t}{r\over \delta}\left[a\,\delta\,g'(a\,\delta)h(r) +
{2\over3}\,a\,\delta\,g'(a\,\delta)\,f(r) + {2\over3}(a\,\delta)^2 g''(a\,\delta)\,f(r)\right]\,,\\
&&\hspace*{-1.5cm}
\ddot A(r,t) = -{F(r)\over2A^2(r,t)}\,,\\
&&\hspace*{-1.5cm}
\ddot A'(r,t) = {F(r)A'(r,t)\over A^3(r,t)} - {F'(r)\over 2A^2(r,t)}\,.
\end{eqnarray}

These functions allow us to construct any other observable. For instance, the
transverse and longitudinal rates of expansion can be written as
\begin{eqnarray}
&&H_T(r,t) \equiv {\dot A(r,t)\over A(r,t)}\,,\\
&&H_L(r,t) \equiv {\dot A'(r,t)\over A'(r,t)}\,.
\end{eqnarray}
Note that in general these two functions will be different, and they enter
into other observables. We can also construct quantities like the
``effective'' acceleration parameter 
\begin{equation}
q(z) = -1 + {d\,\ln\,H(z)\over d\,\ln(1+z)}\,,
\end{equation}
where $H(z)$ is in fact $H_L(r(z),t(z))$.

One could also define an effective equation of state parameter
\begin{equation}\label{wpr}
w(z) \equiv {p(z)\over \rho(z)} = -1 + {1\over3}
{d\,\ln\Big[H^2(z)/H_0^2(r) - \Omega_M(r)(1+z)^3\Big]
\over d\,\ln(1+z)}\,,
\end{equation}
where $H(z)$ is here $H_T(r(z),t(z))$.

\subsection{Parametric solution}
At fixed $r$ the $r$-dependent Hubble rate
Eq.~(\ref{eq:hubblerate}) is just like the normal Friedmann
equation, and the standard way to explicitly solve
for $A(r,t)$ is to use an additional parameter $\eta$.
With the selected gauge the solution is
\begin{equation}\label{eq:Aeta}
A(r,t) = \frac{\Omega_M(r)}{2[1-\Omega_M(r)]} [\cosh(\eta) - 1] A_0(r)
\end{equation}
\begin{equation}\label{eq:timeeta}
H_0(r) t = \frac{\Omega_M(r)}{2[1-\Omega_M(r)]^{3/2}} [\sinh(\eta) - \eta]
\end{equation}
Given $r$ and $t$, by solving Eq.~(\ref{eq:timeeta}) $\eta$
can be found, and combining Eqs.~(\ref{eq:Aeta})-(\ref{eq:timeeta})
with the Einstein equations we can then derive any necessary
quantity.\\[1ex]

We have used both the series solution, implemented in a {\tt Mathematica}
notebook, and the parametric solution, implemented as a Fortran program,
to make and double check all numerical computations in this paper.
The Fortran 90 program together with a set of {\tt IDL} routines for making
likelihood plots is made publicly available and can be downloaded
at \url{http://www.phys.au.dk/~haugboel/software.shtml}.

\subsection{The GBH model}

Here we define a new type of LTB model, which is completely specified by
the matter content $\Omega_M(r)$ and the rate of expansion $H_0(r)$,
\begin{eqnarray}
&&\Omega_M(r) = \Omega_{\rm out} + \Big(\Omega_{\rm in} - 
\Omega_{\rm out}\Big)
\left({1 - \tanh[(r - r_0)/2\Delta r]\over1 + \tanh[r_0/2\Delta r]}\right) \,,\\
&&H_0(r) =  H_{\rm out} + \Big(H_{\rm in} - H_{\rm out}\Big)
\left({1 - \tanh[(r - r_0)/2\Delta r]\over1 + \tanh[r_0/2\Delta r]}\right)  \,,
\end{eqnarray}
which is governed by 6 parameters,
\begin{eqnarray}
&\Omega_{\rm out}\hspace{1cm}& {\rm determined\ by\ asymptotic\ flatness}\\
&\Omega_{\rm in} \hspace{1cm}& {\rm determined\ by\ LSS\ observations}\\
&H_{\rm out}\hspace{1cm}& {\rm determined\ by\ CMB\ observations}\\
&H_{\rm in}\hspace{1cm}& {\rm determined\ by\ HST\ observations}\\
&r_0\hspace{1cm}& {\rm characterises\ the\ size\ of\ the\ void}\\
&\Delta r\hspace{1cm}& {\rm characterises\ the\ transition\ to\ uniformity}
\end{eqnarray}

\begin{figure}
\begin{center}
\includegraphics[width=0.45 \textwidth]{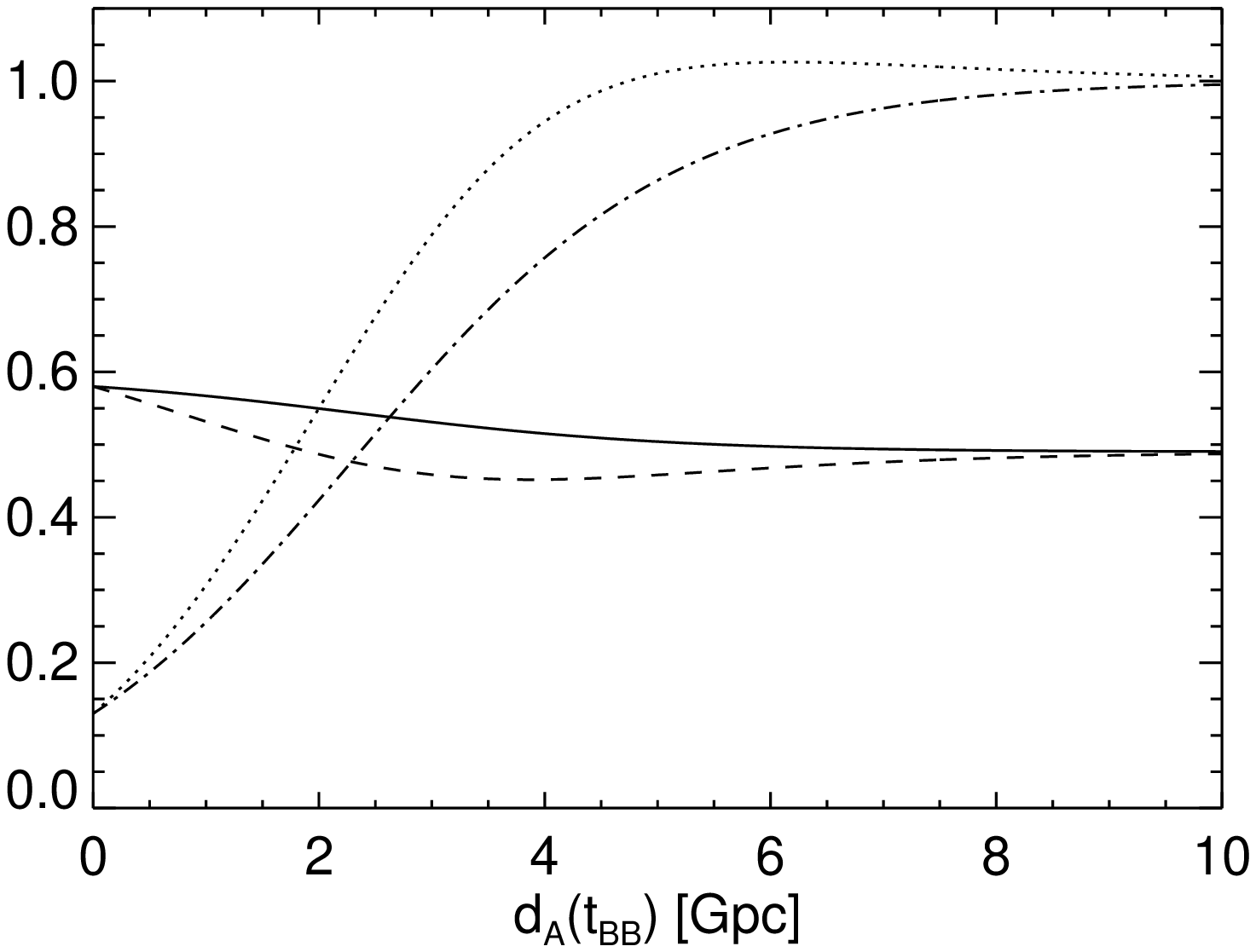}
\includegraphics[width=0.45 \textwidth]{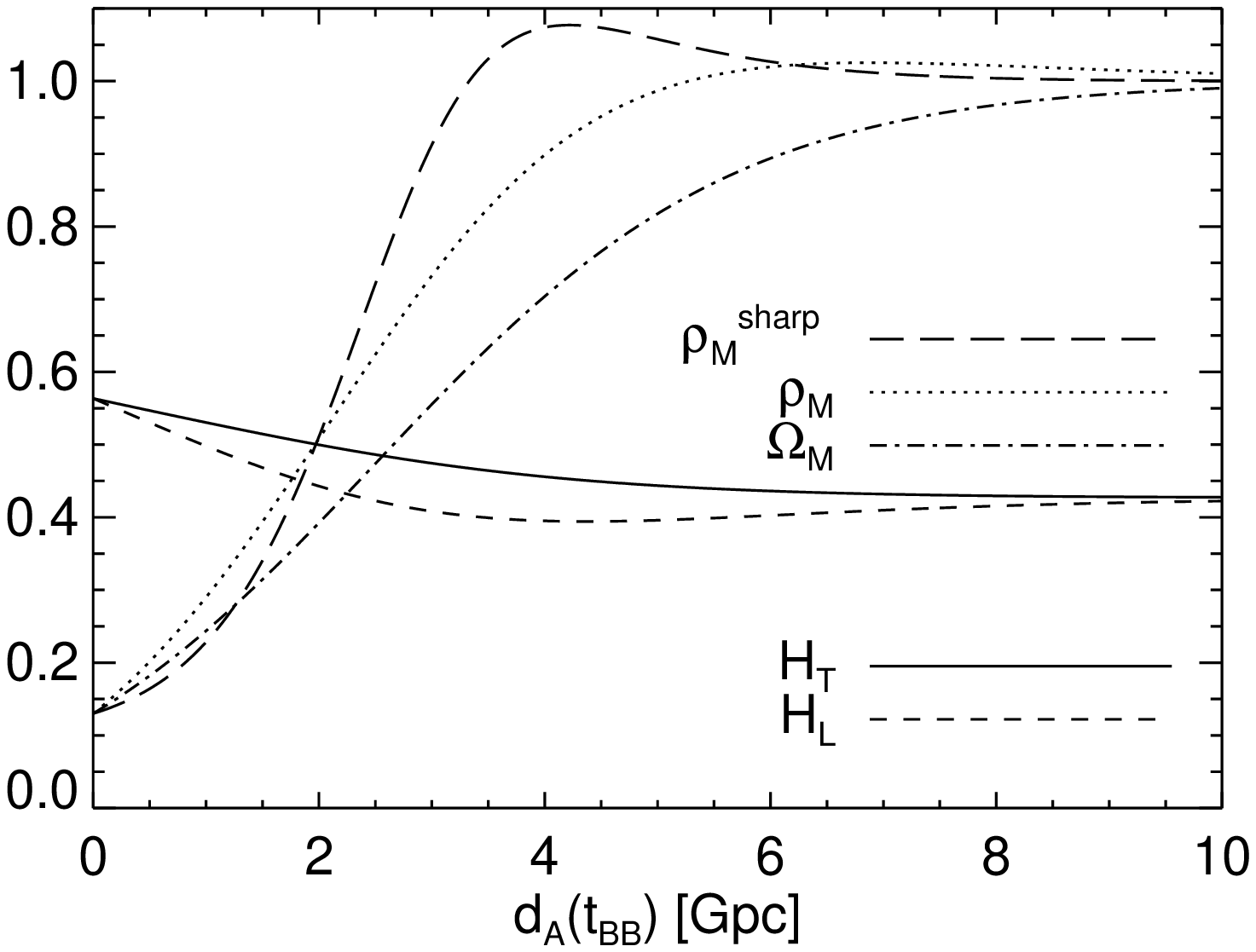}
\caption{The radial dependence of the physical matter density in units of the
critical density ($\rho_m$), of our density function ($\Omega_m$), that coincides
with the usual matter density at the centre and asymptotically, and of the
transverse and longitudinal expansion rates ($H_T$ and $H_L$).
The radial axis is the angular diameter distance ($d_A(r,t_{\rm BB})$), and
everything is taken at the current time for the central observer.
To the left (right) is shown the best fit GBH (constrained)
model (see table \ref{tab:models}). Also shown to the right is an example of the density profile of 
a model with a sharp transition ($\Delta r / r_0 = 0.3$), but still
within 1-$\sigma$ of the best fit.}\label{fig:HOM}
\end{center}
\end{figure}

We fix $\Omega_{\rm out} = 1$ and let the other five parameters vary
freely in our parameter scans (see table \ref{tab:priors} for the priors). For
instance, a plot of $\Omega_M(r)$, $H_T(r)$ and $H_L(r)$ for the two best
fit models, as a function of the angular diameter distance today,
$d_A(r,t_{\rm BB})$, can be seen in Fig.~\ref{fig:HOM}\footnote{Note that while
apparently this model is similar to that of Ref.~\cite{Alnes:2005rw}, it differs
in the details. They fix their gauge $A(r,t)=A_0(r)=r$ at the moment of
recombination, while we fix it at the current time, and they define their
free functions in terms of $F(r)$ and $k(r)$, while we use $H(r)$
and $\Omega(r)$.}. Also shown is the density profile of a model with a
sharper transition ($\Delta r/r_0=0.3$), but still within 1-$\sigma$ of the best
fit. This illustrates that observations allow for shallower density profiles
close to the origin, and that for the physical matter density $\rho_M$, we
naturally get a shell-like transition.

\subsection{The constrained GBH model}

We have also considered a more constrained model, in which the Big Bang
is homogeneous, that is, the spatial hypersurface at the Big Bang does
not depend on the radial coordinate $r$. This can be obtained simply
using Eq.~(\ref{TBB2}), by a choice of $H_0(r)$,
\begin{equation}
H_0(r) = H_0\left[{1\over \Omega_K(r)} -
{\Omega_M(r)\over\sqrt{\Omega_K^3(r)}}\ {\rm sinh}^{-1}
\sqrt{\Omega_K(r)\over\Omega_M(r)}\right]\,,
\end{equation}
so that $t_{\rm BB} = c\,H_0^{-1}$ is universal, for all observers, irrespective
of their spatial location. Note that in this model we have less freedom than
in the previous model, since now there is only one arbitrary function,
$\Omega_M(r)$, and there is one free parameter less. The model is not only
appealing because there are fewer degrees of freedom: only in a
model with a homogeneous time to Big Bang the void corresponds to a perturbation
in a FRW model at early times with no decaying mode \cite{Hellaby:2005}. Ofcourse,
if the void is not a very rare fluctuation in the density field, but created through
a causal process in the early universe, this may not neccesarily be a constraint.

\subsection{Apparent acceleration in the light cone}

It is yet not clear, if what we observe as a function of redshift, in
the form of luminosity distances to standard candles (e.g.~Supernovae
Type Ia), angular diameter distances to standard rulers
(e.g.~Baryon Acoustic Oscillations), the galaxy power spectrum,
galaxy cluster counts, or other measures of the geometry and
mass distribution of the universe, are due to modifications of gravity,
an extra energy component, a cosmological constant, or simply
a wrong interpretation of the underlying cosmological model.
But as a whole they represent different possible explanations for
the ``Dark Energy'' problem. One of the main observables, that
will help decide between the different scenarios in the future, is
the Hubble parameter $H(z)$. {\em Under the assumption}
that the correct background is a flat FRW cosmology we can
write it \cite{Linder:2003dr}
\begin{equation}\label{eq:w}
\hspace{-1.5cm}
\frac{H_{T,L}^2(z)}{H_{\rm in}^2} =  (1+z)^3 \Omega_{\rm in} + (1-\Omega_{\rm in})
\exp\left[ 3 \int_1^{1+z} d\log(1+z') (1 + w^{T,L}_{\rm eff}(z'))\right]\,,
\end{equation}
where $H_{\rm in}$ and $\Omega_{\rm in}$ are the expansion rate and
matter density as observed at $z=0$\footnote{Alternatively,
in \cite{Linder:2005in} Eq.~(\ref{eq:wb}) is written with derivatives
of the scale factor $a$ instead of the redshift $z$.
We advocate using $z$, as it is an observable, in
contrast to $a(z)$. Notice that $H_T(0,t)=H_L(0,t)$.}.
By taking the derivative we can write it as
\begin{equation}\label{eq:wb}
w^{T,L}_{\rm eff}(z) = -1 + \frac{1}{3}\frac{d \log\left[ \frac{H_{T,L}^2(z)}{H_{\rm in}^2} -
         (1+z)^3 \Omega_{\rm in} \right]}{d\log[1+z]}\,,
\end{equation}
where we have assumed that ${H_{T,L}^2(z) \over H_{\rm in}^2} -
(1+z)^3 \Omega_{\rm in} > 0$. The beauty of Eq.~(\ref{eq:wb}) is that if the
observational data indeed is a manifestation of extra energy
components, $w^{T,L}_{\rm eff}(z)$ has the usual interpretation of a dark energy
equation of state, while in the case of modified gravity models, or
the LTB model that we are considering in this paper, it can be
interpreted as an empirical observational signature. In those case
$w^{T,L}_{\rm eff}(z)$ is a function that captures the difference between
the expansion rate that we measure, and the expansion rate that we ascribe to the
{\em observed} matter density $\Omega_m$ (see also Eq.~\ref{wpr}
for a correct definition of $w$ in the case of an LTB universe).

\begin{figure}
\begin{center}
\includegraphics[width=0.95 \textwidth]{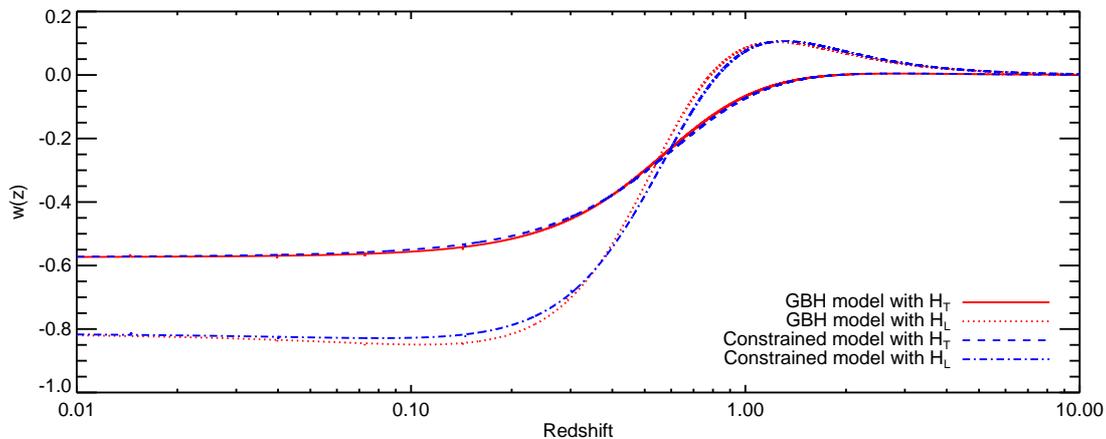}
\caption{The apparent acceleration $w^{T,L}_{\rm eff}(z)$ inferred by applying
Eq.~(\ref{eq:wb}) and FRW cosmology as the underlying model to
describe the change in the Hubble parameter (either $H_T(z)$ or
$H_L(z)$) in the two best fit LTB models.}\label{fig:w}
\end{center}
\end{figure}

It is worthwhile pointing out, that even if there is not an accelerated
expansion in the LTB models we are considering, because data is
observed in the light cone, the total time derivative is \cite{Enqvist:2007vb}
\begin{equation}
{D \over Dt} = {\partial \over \partial t} - {c\sqrt{1 - k(r)}\over A'(r,t)} 
{\partial \over \partial r} \simeq {\partial \over \partial t} - c
{\partial \over \partial r}\,,
\end{equation}
and an observer can measure an {\em apparent} acceleration when the
light cone traverses the central inhomogeneity due to spatial
gradients in either the matter density or the expansion rate. In
Fig.~\ref{fig:w} we show $w^{T,L}_{\rm eff}(z)$ for two best fit models.
Letting $H = H_T$ and $H = H_L$ are both relevant, because different
observations probe different expansion rates, i.e.~as will be seen
below, the Baryon Acoustic Oscillation signal depends partly on $H_L$,
while supernova observations are only related to $H_T$ through its
dependence on $d_L = (1+z)^2 d_A = (1+z)^2 \exp(\int H_T dt)$.
Interestingly, the variation and derivative of $w^{T,L}_{\rm eff}(z)$ is 
quite large in the best fit LTB models, showing that a precise low redshift
supernova survey, such as the SDSS Supernova Survey~\cite{SDSS-SN},
sensitive to $H_T$, or a fine grained BAO survey such as PAU~\cite{PAU},
sensitive to $H_L$, could rule out or reinforce the models in the near future.
Conversely, if a disagreement between $w$ as observed by Supernovae and
$w$ as observed through the BAOs is found, this could be a hint of
inhomogeneous expansion rates.

We can directly compute $w^{T,L}_{\rm eff}(z)$ in
the limiting cases $z=0$ and $z \gg 1$ for asymptotically flat LTB
models
\begin{equation}
w^{T,L}_{\rm eff}(z) = \left\{ \begin{array}{ll}
- \frac{1}{3} + \frac{2}{3}\frac{c H'_0(0)}{(1 - \Omega_{\rm in})H_{\rm in}^2}
& \textrm{if $z=0$, and $H=H_T$}\\
- \frac{1}{3} + \frac{4}{3}\frac{c H'_0(0)}{(1 - \Omega_{\rm in})H_{\rm in}^2}
& \textrm{if $z=0$, and $H=H_L$}\\
0 & \textrm{if $z\gg1$}\end{array} \right.\,,
\end{equation}
where we have used that the LTB metric converges asymptotically
to a FRW metric giving $dz = - da / a$. We see that to have
$w \ll -1/3$ at low $z$ implies either a significant negative
gradient in $H_0(r)$, or $\Omega_{\rm in} \sim 1$.

\subsection{Physical scales in the early universe}\label{sec:scale}

Many bounds from observational cosmology, such as the sound horizon
(a ``standard ruler''), the CMB, and the big bang nucleosynthesis, are derived
by considering scales and processes in the early universe, and are based
on the implicit assumption of an
underlying FRW metric. To test LTB models against these observational
data we have to connect distance scales, redshifts, and expansion rates
in the early universe to those observed today.

By construction, at high redshifts the LTB metric converges to a FRW metric,
and the central void disappear (see Eq.~\ref{eq:hubblerate}), and physical
results derived for FRW space-times still hold in the early universe,
even though we are considering an LTB space-time.
But starting from an approximately uniform universe at a high redshift
$z_e$ in the LTB model, the expansion rate and matter density become
gradually inhomogeneous, and a uniform comoving physical scale
$l$ in the early universe at $z_e$, for example the sound horizon, is
not uniform at some later redshift $z$. In particular the comoving size at
$t = t_0 = t_{\rm BB}(0)$ depends on how much relative expansion there
has been at different positions since the formation of the uniform scale
\begin{equation}
l(r(z)) = l(r_\infty) \frac{A(r(z),t_0)}{A(r(z),t(z_e))}
\frac{A(r_\infty,t(z_e))}{A(r_\infty,t_0)}\,,
\end{equation}
where $t_0$ is the time now for the central observer,
and $r_\infty$ is the radial coordinate of an observer very far away 
from the void. This is a consequence of defining the comoving physical
scale as the scale measured at $t_0$. If instead we fixed the comoving
length scales to be measured in the early universe at $t(z_e)$, then
indeed $l(r(z))$ would be independent of the observer position.
The convenience of the above formula is, that the LTB models we
consider are asymptotically FRW, and we can easily compute
comoving scales at infinity.

Normal relations determining early universe quantities, such as the
redshift at matter-radiation equality, are essentially based on the
Friedmann and conservation equations to relate cosmological
parameters now to the parameters then
\begin{eqnarray}\label{eq:friedmann}
H^2(z) &= H_0^2\left[\Omega_M(1 + z)^3 + (1-\Omega_M)(1+z)^2\right] \\
\rho(z) a^3 &= \rho(0) a_0^3\,,
\end{eqnarray}
where we have written it for a matter dominated universe.
Eq.~\eqref{eq:hubblerate} is the LTB equivalent to Eq.~\eqref{eq:friedmann},
and since we are considering LTB metrics that asymptotically converge
to FRW metrics the equation for light rays, or null geodesics, \eqref{eq:null}
at high redshifts has the usual solution $d(1+z) = - da/a$, and we can
write an asymptotic version of Eq.~\eqref{eq:hubblerate} that is valid
at high redshifts 
\begin{equation}
H^2(z) = H_{\rm eff}^2\left[\Omega^{\rm eff}_M(1 + z)^3 +
(1-\Omega^{\rm eff}_M)(1+z)^2\right]\,
\end{equation}
where
\begin{equation}\label{eq:heff}
H_{\rm eff} = H_0(z_e) \left[\frac{A_0(r(z_e))}{(1+z_e) A(r(z_e),t(z_e))}\right]^{3/2}\,
\end{equation}
and the asymptotic matter density is
\begin{equation}\label{eq:omegaeff}
\Omega^{\rm eff}_M = \Omega_M(r(z_e)) = \Omega_{\rm out} = 1
\end{equation}

In summary, any quantity at high redshifts in the LTB model can be computed
with the usual formulas valid for a FRW metric, but using the matter density
and Hubble constant given in Eqs.~\eqref{eq:heff}-\eqref{eq:omegaeff}.

\section{Observational Data}

To assess the viability of the proposed models we have tested them
against a set of current observational data. We divide the data into
two classes: Constraining data sets, and prior data. The constraining
data sets are actual measurements with errors, that are used to
compute the likelihood of a given model, while the prior data, merely
give ranges inside which the models should be.

\subsection{The Cosmic Microwave Background}

It is not a priori clear how to compute the spectrum of temperature
anisotropies, without a full perturbation theory for the LTB models,
but as shown in section \ref{sec:scale} we can compute
the comoving distance to the surface of large scattering, and the
comoving size of the sound horizon at large distances. The ratio give
the typical size, of the CMB temperature fluctuations on the sky, or
equivalently the scale of the first peak, which is measured with
exquisite precision by the WMAP satellite \cite{WMAP3}
\begin{equation}
\theta_{\rm CMB} = \frac{r_s(\zrec)}{d_C(\zrec)} 
 = 0.5952\pm0.0021^\circ  = 0.010388 \pm 0.00037.
\end{equation}
The sound horizon $r_s(\zrec)$ is calculated using the fitting
formula provided by Eisenstein and Hu \cite{Eisenstein:1997ik} with
$\Omega_M = \Omega_{\rm out} = 1$, $h=H_{\rm eff}$ (see
Eq.~\ref{eq:heff}). The physical baryon density in the early
universe and the recombination redshift are fixed to their best fit
WMAP3 values $\Omega_B H_{\rm eff}^2 = 0.0223$ and
$\zrec=1089$. The $\chi^2$ from the CMB constraint is
simply
\begin{equation}
\chi^2_{\rm CMB} = \frac{\left[\theta_{\rm CMB} - 
     r_s(\zrec)/(d_C(\zrec)\right]^2}
{\sigma_\theta^2}
\end{equation}

\subsection{Baryon Acoustic Oscillations}

The BAO has been measured at different scales using a variety of
techniques, and the feature has been detected in the 3D two-point
correlation function \cite{Eisenstein:2005su,Okumura:2007br}, the 3D
power spectrum \cite{Percival:2007yw}, and the angular power spectrum
\cite{Padmanabhan:2006ku}. In particular Percival et al
\cite{Percival:2007yw} have combined the 2DF and SDSS large scale
surveys to yield a measure of the BAO centred at two different
redshifts, namely $z=0.2$, and $z=0.35$.

\begin{figure}
\begin{center}
\includegraphics[width=0.45 \textwidth]{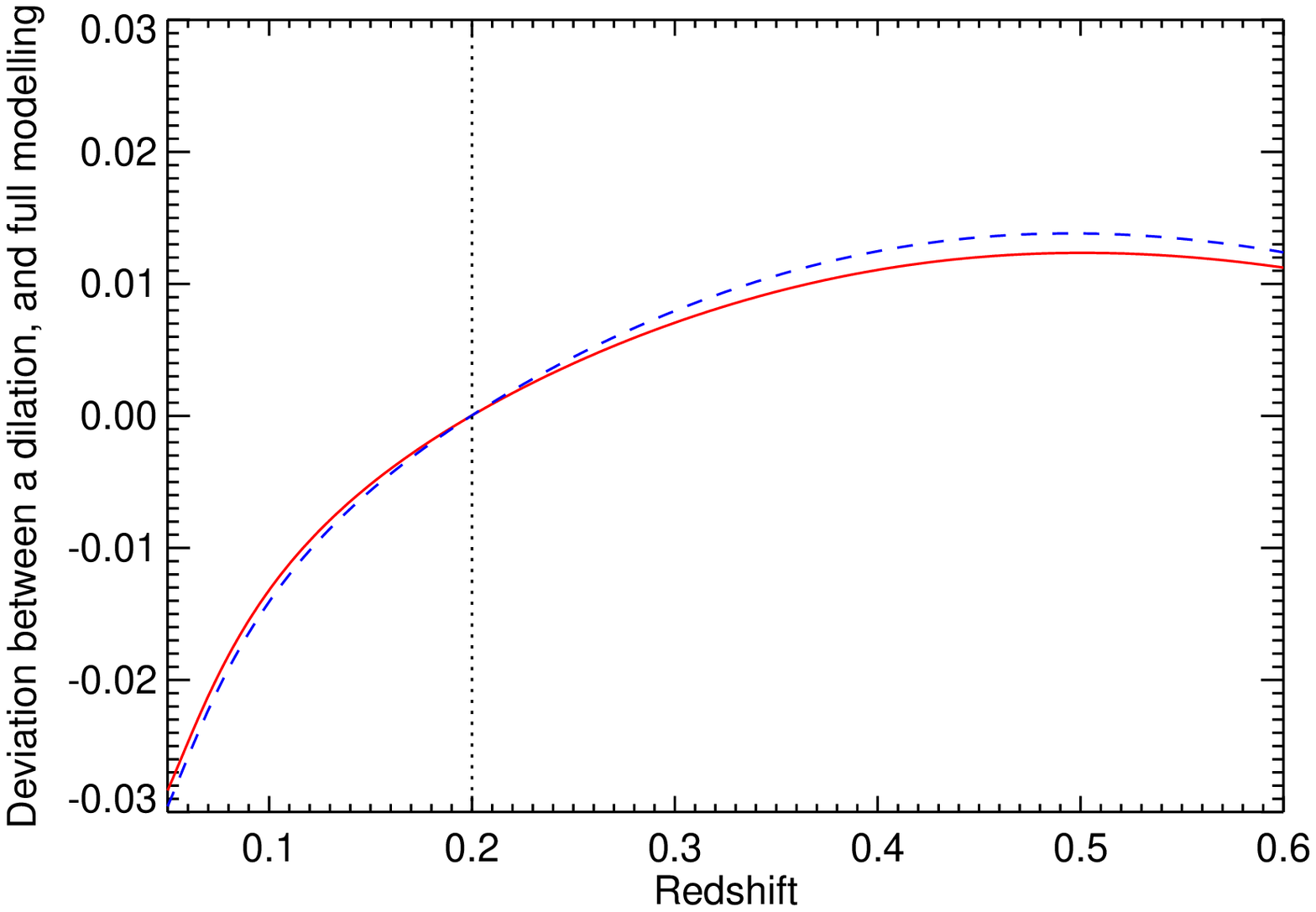}
\includegraphics[width=0.45 \textwidth]{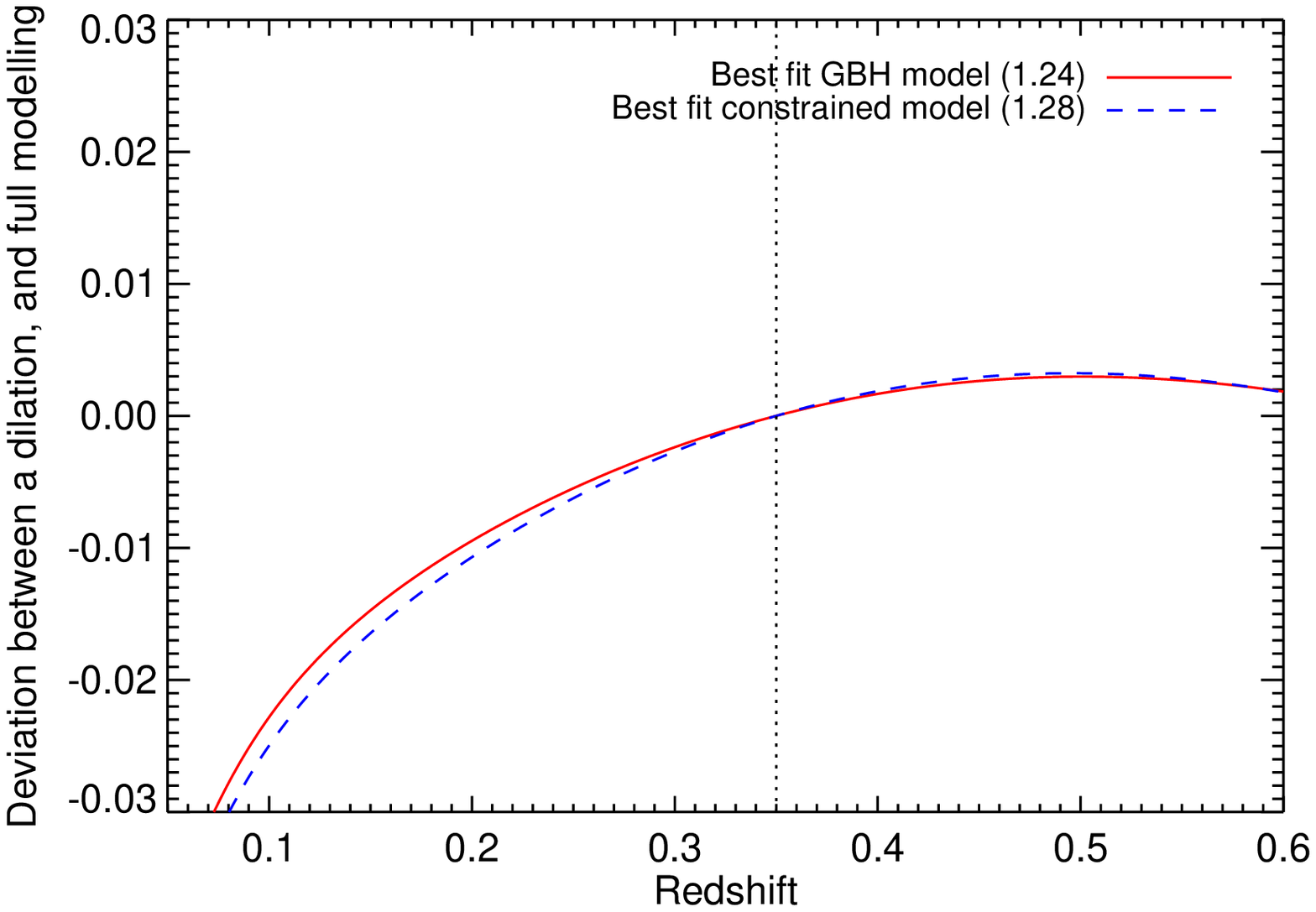}
\caption{The fractional difference in the comoving distance used in
the reference model in \cite{Percival:2007yw} compared to the best fit GBH
model (red, full line) and constrained model (blue, dashed line).
To the left (right) is shown the the dilation at $z=0.2$ ($z=0.35$),
and in parenthesis in the legend is indicated the
dilation.}\label{fig:dilation}
\end{center}
\end{figure}

In \cite{Percival:2007yw} the power spectrum is calculated in a
reference cosmology, and then the comoving distance scale is either
dilated, using a fixed factor, or deformed using a 3 node spline fit.
In principle one would need a full perturbation theory for LTB
geometries to recalculate the 3D power spectrum, using our best fit
model, but if we take into account that the observed galaxies are
divided into redshift slices, and that the LTB universe at constant
redshift behaves locally like a homogeneous FRW universe
(e.g.~Eq.~(\ref{eq:hubblerate}) is a local analogy to the normal FRW
Hubble rate equation), we can then, as a first approximation, relate
the $\Lambda$CDM power spectrum calculated in \cite{Percival:2007yw}
to the LTB power spectrum through a dilation. In
Fig.~\ref{fig:dilation} we show the fractional difference between a
simple dilation of the comoving scale, and a full modelling around the
relevant redshifts $z=0.2$ and $=0.35$. The dilation is an excellent
approximation, the difference being less than 2\% over the relevant
redshift range. We stress, however, that a comprehensive test of LTB
models against large scale structure data has to await the development
of the linear perturbation theory for LTB space-times, an approach
that is outside the scope of this paper~\cite{Alicia}.

The relevant BAO quantity to test against is the observed peak (in the
case of real space) or wiggle wavelength (in the case of Fourier
space), compared to the size of the sound horizon in the reference
model used in \cite{Percival:2007yw}, which can be interpreted approximately as
the projected size of the BAO on the sky for structures at a given
redshift. The ratio is given as
\begin{equation}
\theta_{BAO}(z) = \frac{r_s(z)}{D_V(z)}\,.
\end{equation}

As discussed in Section \ref{sec:scale}, the \emph{comoving} size of
the sound horizon $r_s(z)$ is a function of redshift, due to the
inhomogeneous nature of the expansion since the surface of last
scattering $A(r(z),t(z))/A(r(\zrec),t(\zrec))$. Because we
detect the BAO in the 3D distribution of galaxies, and not in the 2D
projection on the sky, $D_V$ is not just the comoving distance, but
rather a combination of longitudinal and transversal distances. In LTB
space-times the longitudinal and transversal Hubble parameters, $H_L$
and $H_T$, can differ significantly, see Fig.~1, and therefore it is
important to use the correct longitudinal expansion rate:
\begin{equation}
D_V(z) = \left[ d_A^2(z) (1+z)^2 \frac{c z}{H_L(z)}\right]^{1/3}\!.
\end{equation}

Percival et al \cite{Percival:2007yw} find
\begin{equation}
\theta_{\rm BAO}(0.2)   = 0.1980\pm0.0058 \qquad
\theta_{\rm BAO}(0.35) = 0.1094\pm0.0033
\end{equation}
with a 39\% correlation, and we use these two
measurements as our BAO data set giving the $\chi^2$
\begin{equation}
\chi^2_{\rm BAO} = \sum_{i,j}
  \left[\theta_{\rm BAO}(z_i)-r_s(z_i)/D_V(z_i)\right] C_{ij}^{-1}
  \left[\theta_{\rm BAO}(z_j)-r_s(z_j)/D_V(z_j)\right]
\end{equation}
with
\begin{equation}
\boldsymbol{C}^{-1} = \left(\begin{array}{cc}
35059  & -24031 \\
-24031 & 108300
\end{array}\right)
\end{equation}

\subsection{Type Ia Supernovae}

We use the Type Ia Supernovae compilation by Davis et
al.~\cite{Davis:2007na}, which is a compilation of 192 SNIa consisting
of 45 SNe from a nearby SNIa sample~\cite{Jha:2006fm}, 57
SNLS~\cite{Astier:2005qq} and 60 ESSENCE~\cite{WoodVasey:2007jb}
intermediate redshift SNe and 30 high redshift ``Gold''
SNe~\cite{Riess:2006fw}, with internally consistent magnitude
offsets. The supernovae span the redshift range $z=0.01-1.7$, and we
use the magnitude residuals $\mu$ to constrain the LTB models.  The
residual $\mu$, and apparent and absolute magnitudes $m$ and $M$ are
related to the luminosity distance $d_L$ as
\begin{equation}
\mu = m - M = 5 \log_{10} \left[\frac{d_L}{1 Mpc}\right] + 25
\end{equation}
The exact absolute magnitude of a SNIa is unknown, and we include an
arbitrary offset $\mu_0$, when calculating $\chi^2$ for the model fit to
the observed SNIa
\begin{equation}
\chi^2_{\rm SNIa} = \sum_{i}
  \frac{\left[\mu_i^{\rm obs} - (\mu^{\rm model}(z_i) + 
\mu_0)\right]^2}{\sigma_i^2}
\end{equation}
where $\mu_0$ is determined by minimising $\chi^2_{\rm SNIa}$.
$\mu$ has a logarithmic dependence on $d_L$, and the zero point
$\mu_0$ is degenerate with the local overall scale of the expansion rate
($H_{\rm in}$ in the GBH model, and $H_0$ in the
constrained model, see Figs.~\ref{fig:likelihood1} and \ref{fig:likelihood2}).

\begin{figure}
\begin{center}
\includegraphics[width=0.95 \textwidth]{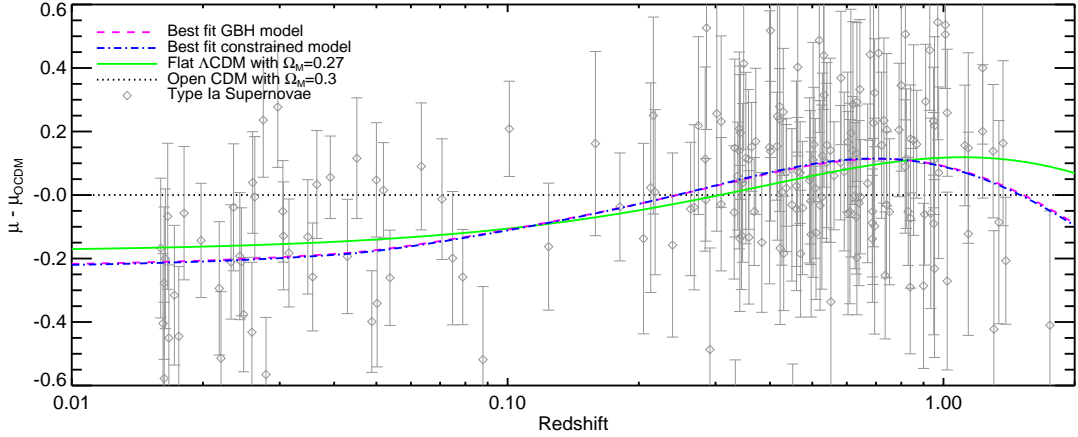}
\caption{Apparent magnitude residuals for the two best fit LTB models,
standard open CDM, and the best fit $\Lambda$CDM  FRW models
compared to Type Ia Supernovae data.}\label{fig:sne}
\end{center}
\end{figure}

Even though the Supernova data set is by far the largest of the three,
the error bars on individual SNe are large, and internally there is a
large scatter, as can be appreciated in Fig.~\ref{fig:sne}, where the 
residuals are compared to the best fit GBH LTB models, and the open 
CDM and $\Lambda$CDM FRW models. Note that the predicted curves for the
best fit GBH model and the best fit $\Lambda$CDM model start to deviate
significantly beyond redshift $z=1$, and therefore it would be extremely 
useful to have a complete Supernovae data set at high redshifts, which
could help discard one of the models against the other one.

The three data sets discussed above are obtained through different
means, and relate to different physical observations, and hence they
are not correlated in any way. This independence allow us to find the
total likelihood by simply multiplying the individual likelihood functions.

\subsection{Priors}

Even though we do not include them in the likelihood analysis, we still require
our models to obey three additional priors: Two of them concern the local universe,
and as such are priors at $z=0$. They are the observed lower age limit on
globular clusters in the Milky Way of 11.2 Gyr \cite{Krauss:2003em}
(2-$\sigma$ limit) and the HST key project \cite{Freedman:2000cf} measure
of the \emph{local} value for the Hubble parameter $H_{\rm in} = 72\pm8$
(1-$\sigma$ limit). The third prior is the gas fraction as observed in clusters of
galaxies. This is a very powerful observation for limiting alternative models:
Clusters of galaxies sit at the bottom of deep gravitational potential wells, and supposedly
the gas fraction is representative for the universe as a whole, because neither
gas nor dark matter can escape out of the potential well. This universal
prior can be compared to the gas fraction we deduce far away from the void,
by combining the WMAP  satellite and Big Bang Nucleosynthesis bound on the
physical baryon density $\omega_b = \Omega_b H_{\rm eff}^{2} = 0.0223$ with
the physical matter density at infinity
$\omega_m = \Omega_{\rm out} H_{\rm eff}^2 = H_{\rm eff}^2$.
The current observational limits are
$f_{\rm gas} = \omega_b / \omega_m = 0.1104\pm0.0016\pm0.1$ (random+systematic)
\cite{Allen:2007ue},
while for our best fit models (see Table \ref{tab:models}) we find
$f_{\rm gas} = 0.127-0.134$ in agreement at 2-$\sigma$ with observations.

\begin{table}[h]
\begin{center}
\begin{tabular}{@{}c|c@{ }c@{ }c|c|c@{ }c}
\hline \hline
Model & $H_0$ & $H_{\rm in}$ & $H_{\rm out}$ & $\Omega_{\rm in}$
           & $r_0$ & $\Delta r$ \cr
{\footnotesize units}
           & \multicolumn{3}{c|}{{\footnotesize 100 km s$^{-1}$ Mpc$^{-1}$}} &
           & {\footnotesize Gpc} & $r_0$ \cr
\hline
GBH            & $-$ & $0.5-0.85$ & $0.30-0.70$ & $0.05-0.35$
                    & $0.3-4.5$ & $0.1-0.9$ \cr
Constrained& $0.50-0.95$ & $0.4-0.89$ & $0.33-0.63$ & $0.05-0.35$
                    & $0.5-4.5$ & $0.1-0.9$ \cr
\hline \hline
\end{tabular} 
\end{center}
\caption{Priors used when scanning the parameters of the two models.
In the constrained model $H_0$ is only a pre factor for $H_0(r)$ and
the span of $H_{\rm in}$ and $H_{\rm out}$ are derived from the
priors on $\Omega_{\rm in}$ and $H_0$.}\label{tab:priors}
\end{table}

\section{Analysis and results}
To test the full and the constrained GBH LTB model, we have
performed a parameter scan over the models and for each set calculated the
$\chi^2$. The priors are given in table \ref{tab:priors}, and have been chosen
to encompass the best fit 2-$\sigma$ limits, except where large degeneracies
exist, and also to be reasonable, taking into account  the HST key project
\cite{Freedman:2000cf}, and acceptable matter densities.

\subsection{The best fit models}
In table \ref{tab:models} are given the best fit for the two models. It is interesting
to notice that both have very similar values, and in fact this seems to hint that
current data prefer the simpler constrained model with a homogeneous big bang.
Both best fit models have local Hubble rates on the low side but still in agreement
with the HST project, and the local time to Big Bang is well inside the limits given by
 globular clusters.
 
The best fit models both give an excellent luminosity redshift distance relation,
that are in as good an accordance with current Type Ia SNe as the
$\Lambda$CDM model (see Fig. \ref{fig:sne}). This comes as no surprise: LTB
models can be constructed that fit {\em any} luminosity-redshift relation
\cite{Celerier:1999hp}.
Our model fit is done though under the simultaneous constraints of the other probes,
and as such is more constrained. In the transition zone between the void and the
surrounding Einstein-de Sitter space (at $r \sim r_0$) there is a significant
(up to $\sim$10\%) difference between $H_T(r)$, which is related to $d_L$,
important for supernova observations, and $H_L(r)$, important for the longitudinal part
of the Baryon Acoustic Oscillations (see Fig.~\ref{fig:HOM}). This difference marks
a fundamental observational signature between LTB and FRW models.
A set of very well observed Type Ia SNe at $z~0.1-0.2$, such as the SDSS II SN survey
\cite{SDSS-SN}, together with the BAO observations already done at similar redshifts,
will put strain on either model. 

\begin{figure}
\begin{center}
\includegraphics[width=0.45 \textwidth]{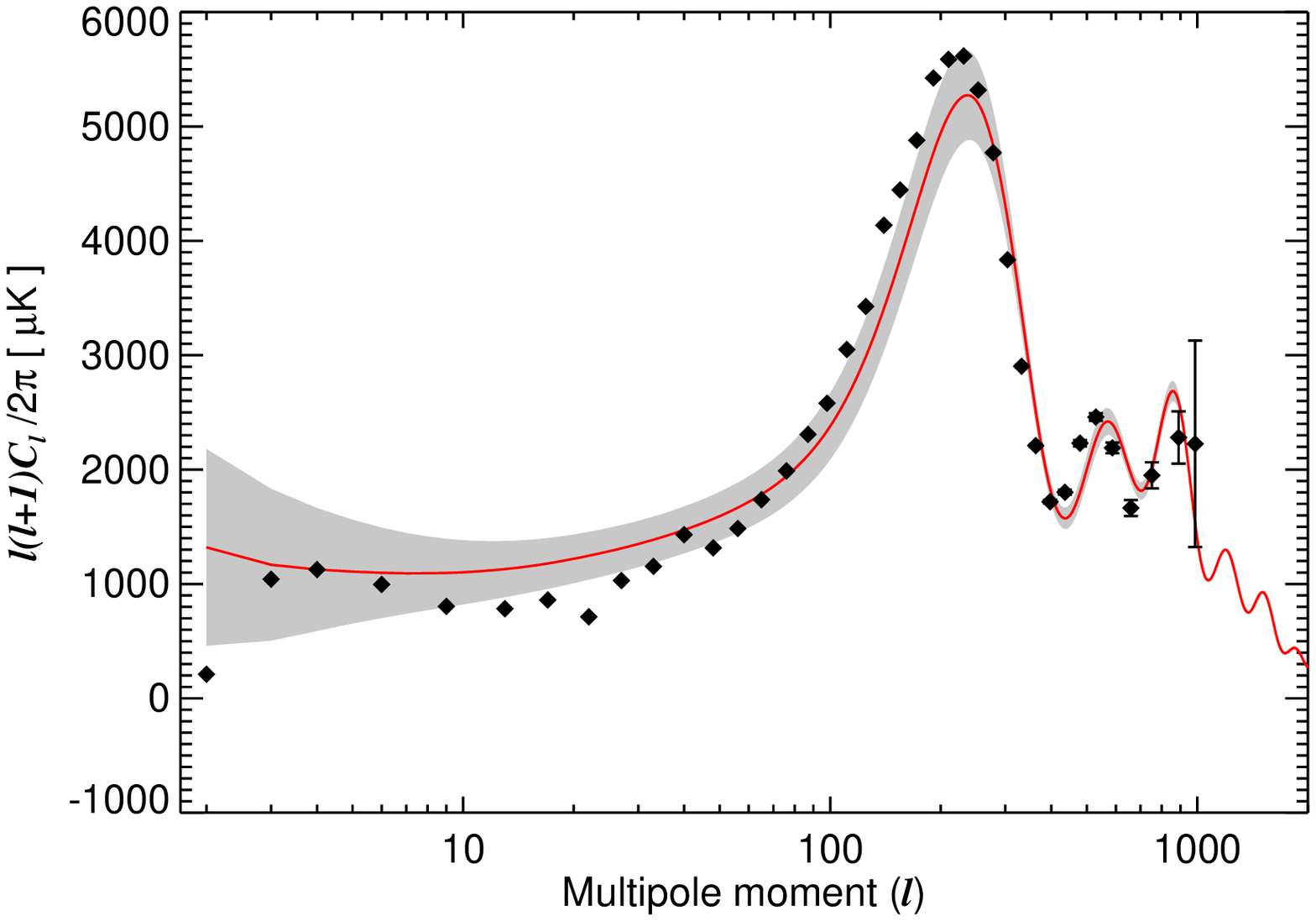}
\includegraphics[width=0.45 \textwidth]{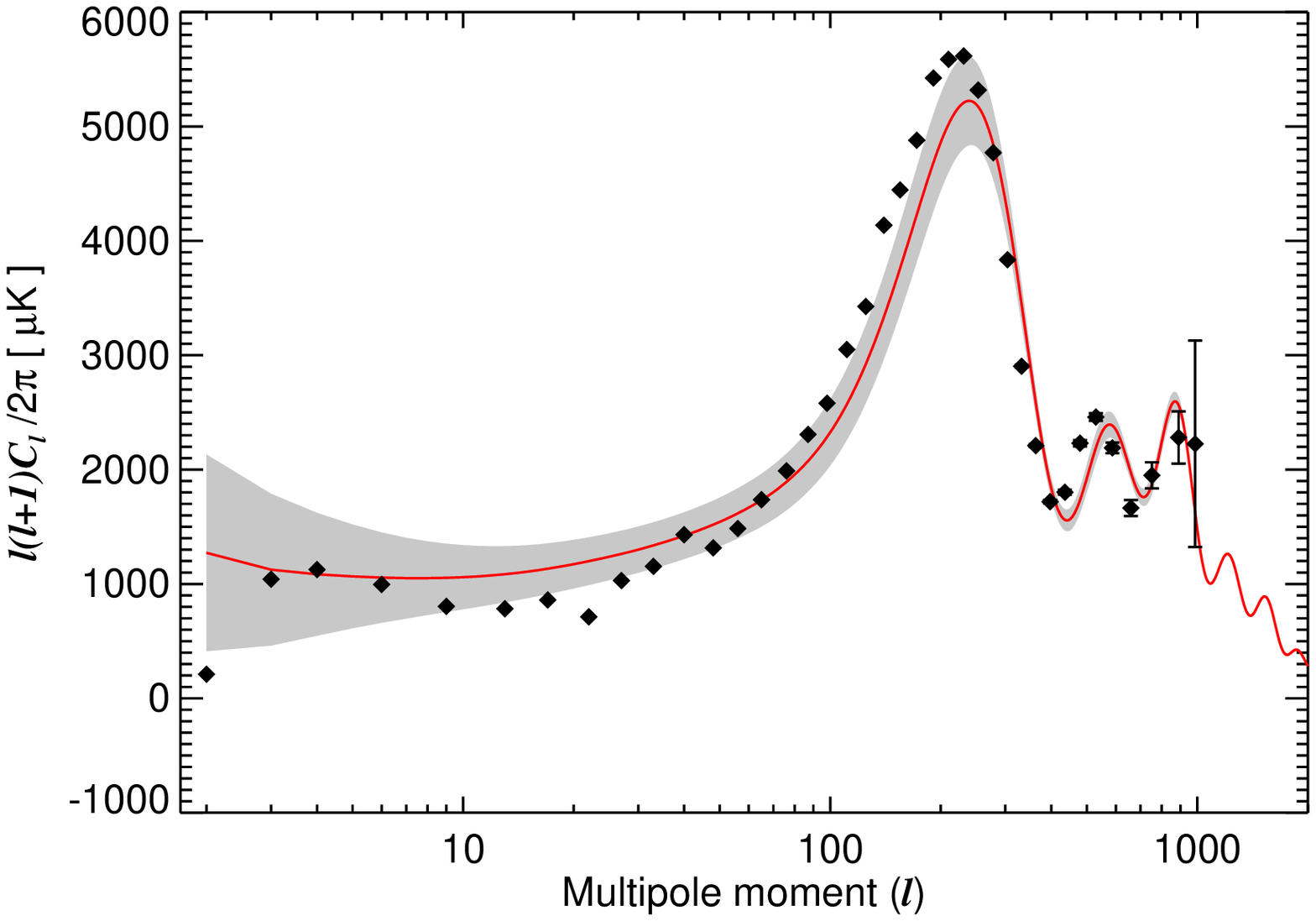}
\caption{To the left/right the CMB spectrum for the best fit 
GBH/constrained-GBH model (red line) compared to WMAP3 data
(diamonds) \cite{Hinshaw:2006ia}, and including cosmic variance
(grey shading).}\label{fig:cmb}
\end{center}
\end{figure}

Because we currently do not have a full perturbation theory for LTB 
space-times, we were not able to make a full likelihood analysis 
comparing our model to all the WMAP data. Nonetheless, to get an idea 
of how bad the models fit the full body of WMAP temperature anisotropy 
observations, we have calculated a standard temperature anisotropy 
spectrum using as input $H_{\rm eff}$ and $\Omega=\Omega_{\rm out}=1$
(see Fig.~\ref{fig:cmb}).
Even though we fix the physical baryon density, and only fit the first peak of the CMB,
the obtained model is not too bad, and it is reassuring that other people have proposed
Einstein-de Sitter models that do fit the WMAP3 CMB observations using a non standard
primordial spectrum and a hot neutrino component~(e.g.~\cite{Hunt:2007dn}).
Very recently an LTB model similar to ours was proposed \cite{Alexander:2007xx} and the
authors managed to make a reasonable fit to the full WMAP TT and TE data by adding
a running of the tilt in the primordial spectrum.
In a more complete analysis at low $l$-values and large angles one could
expect an effect similar to that of a cosmological constant, since we have a non-trivial
curvature at low redshifts, and hence an even better fit that do not have to rely on ad hoc
features or a large running of the tilt in the primordial power spectrum may indeed be possible.

\subsection{Likelihood contours and degeneracies}
\begin{figure}
\begin{center}
\includegraphics[width=0.45 \textwidth]{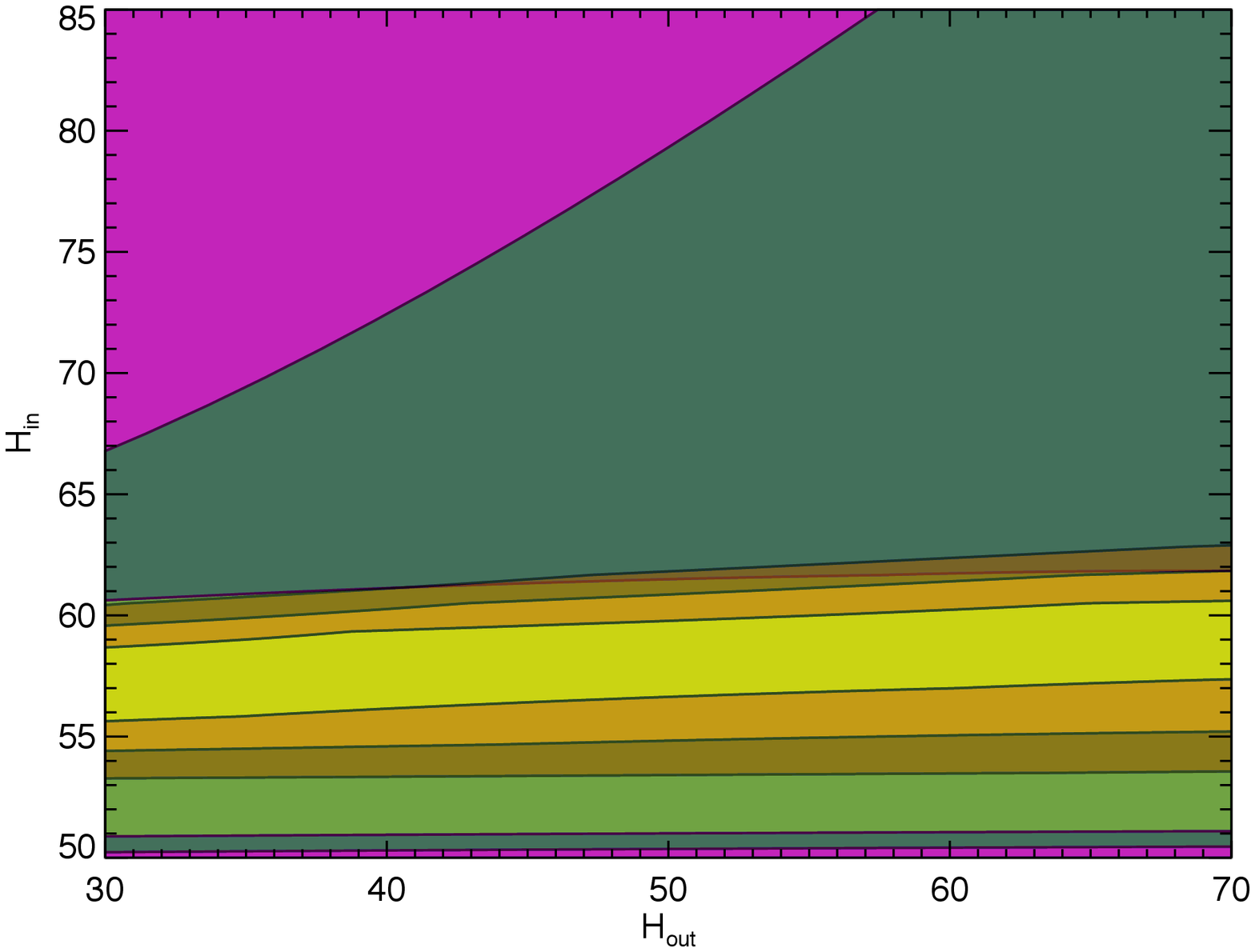}
\includegraphics[width=0.45 \textwidth]{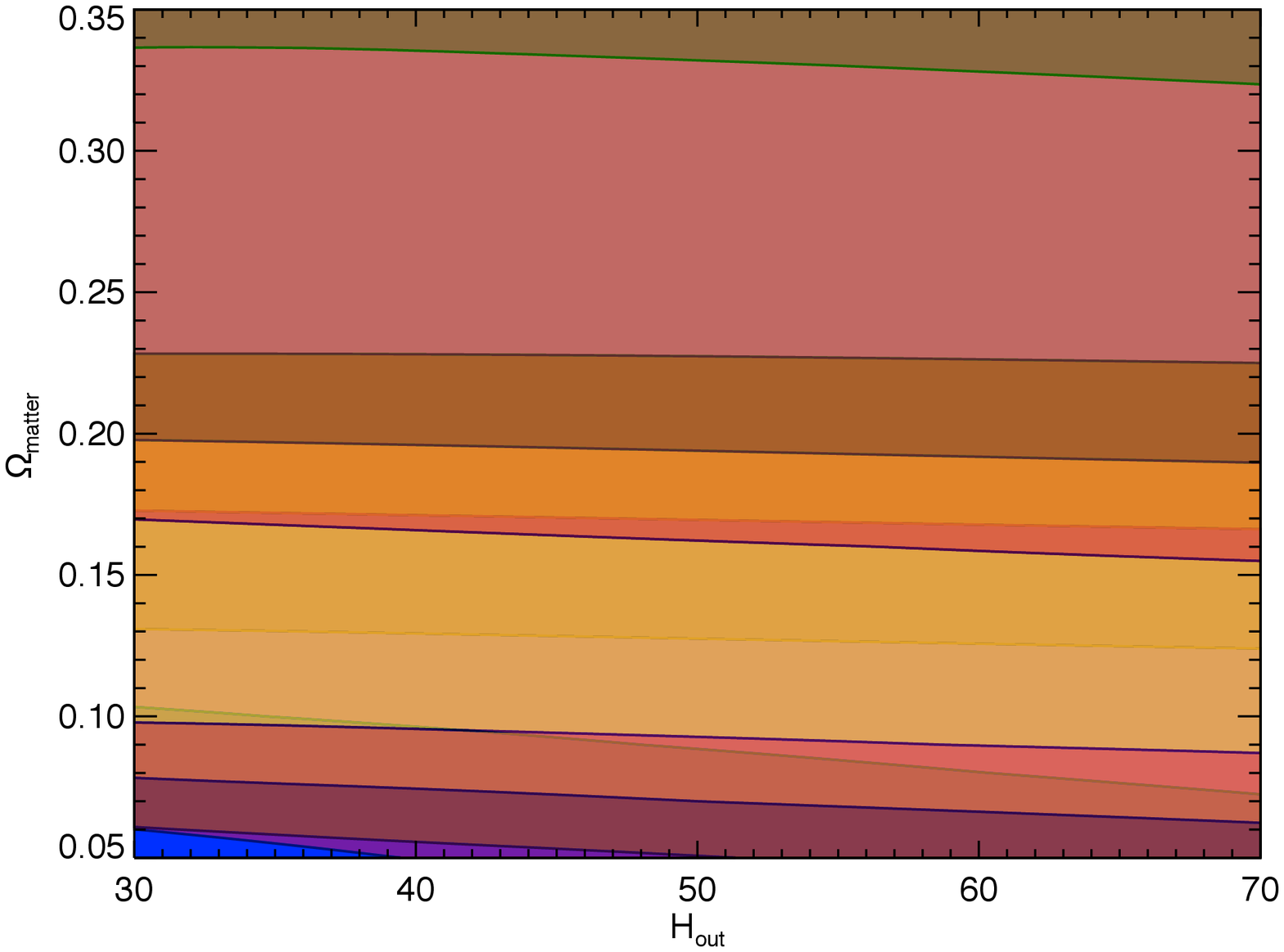}
\includegraphics[width=0.45 \textwidth]{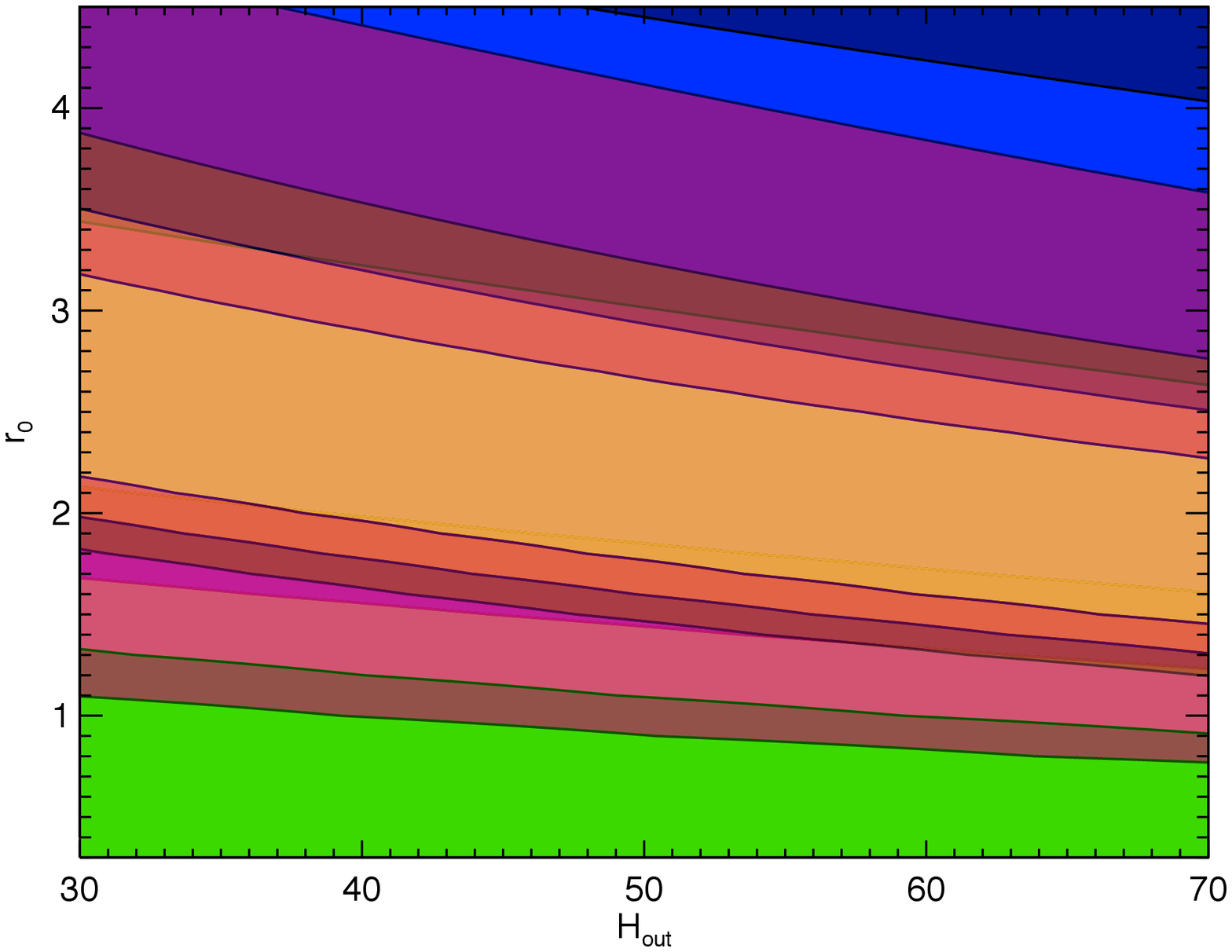}
\includegraphics[width=0.45 \textwidth]{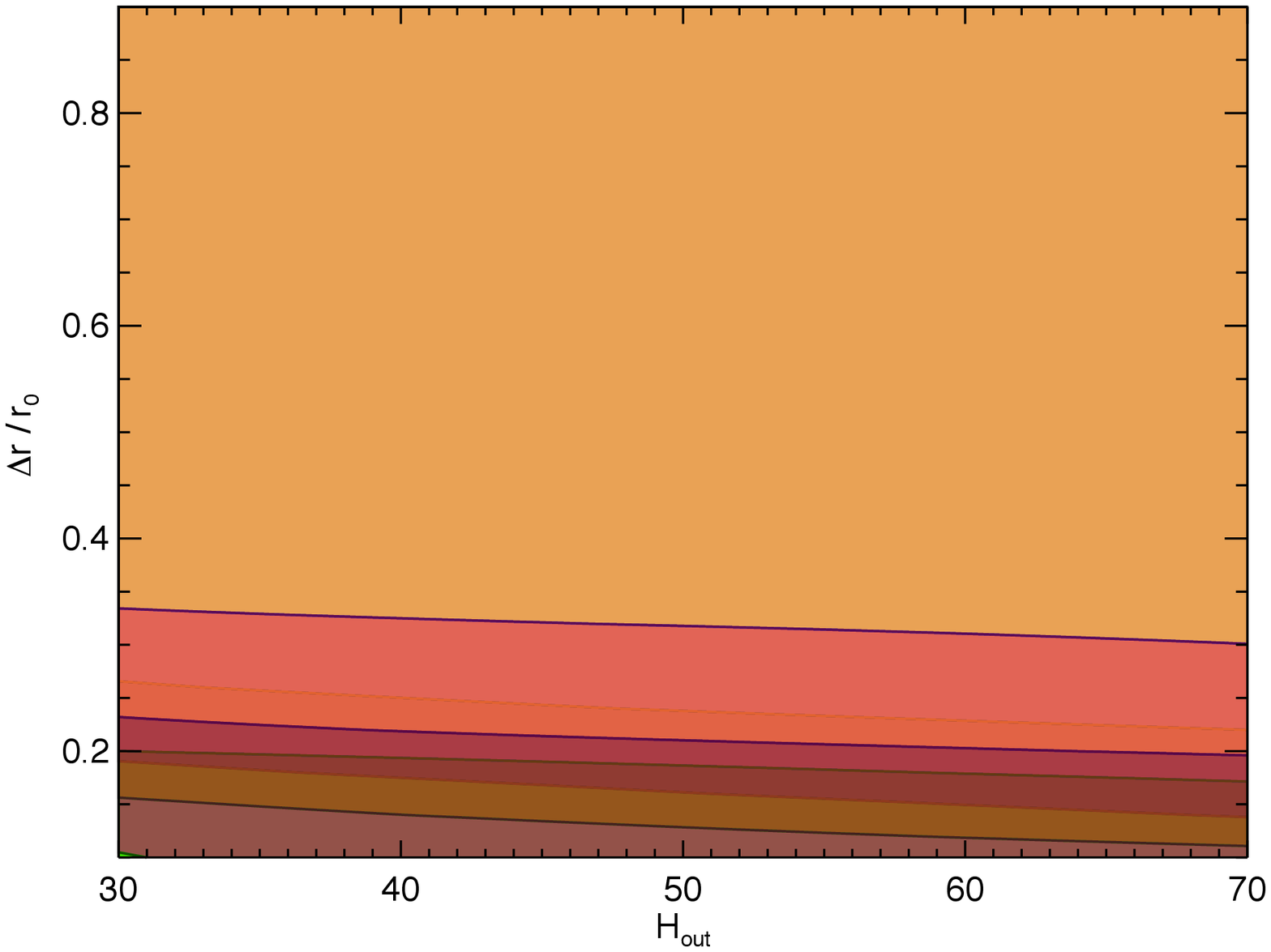}
\caption{Likelihoods for the GBH model: The likelihood for the
combined data set is shown in yellow with 1-, 2-, and 3-$\sigma$
contours, while the individual SNIa, BAO, and CMB data sets are
shown in blue, purple, and green respectively with 1- and
2-$\sigma$ contours.}
\label{fig:likelihood2}
\end{center}
\end{figure}

We have marginalised over different dimensions in parameter space by
integrating over the likelihood, given as $\mathcal{L} \propto \exp(-\chi^2/2)$.
The marginalised 1-$\sigma$ and 2-$\sigma$ likelihoods for the individual
data sets, and also the 3-$\sigma$ limits for the combined data sets are
shown in Figs.~\ref{fig:likelihood1} and \ref{fig:likelihood2}.

In the normal $\Lambda$CDM model, if $\omega_b$ is fixed, there is
a well known strong degeneracy between $\Omega_m$ and
$H_0$\cite{Page:2003} for a given size of the sound horizon on the
sky $\theta_A$. In our model the relevant $\Omega_m$ for the CMB is
$\Omega_{\rm out}=1$, that fixes the size of the sound horizon, and we
have also fixed $\omega_b$. Then the size of $\theta_A$ depends essentially
on $H_0$ or $H_{\rm in}$, which is reflected in the likelihood constraint from
CMB on Hubble rates as seen in the figures. Nonetheless we can to
some extent change $d_A$ by introducing curvature, allowing us to choose
a higher value for $H_{\rm in}$ or $H_0$, either by having a large void size, $r_0$,
or making the void very underdense $\Omega_{\rm in} \ll 1$.
This can be seen by the widening of the 2-$\sigma$ limit in the $H$-$r_0$
plot for large values of $r_0$, and the asymmetric  2-$\sigma$ errors in the
$H$-$\Omega_{\rm in}$ plots.

$\Omega_{\rm in}$ and $r_0$ are the major parameters determining  the
luminosity-redshift relation, and are hence constrained by the Type Ia SNe
and the BAO. While only the relative value of Hubble rate play a role for
the SNe, the BAO does limit $H$, and there is some strain between the
BAO and the Type Ia SNe, as seen in the $\Omega_{\rm in}$-$r_0$ plots.

An obvious degeneracy is that of $r_0$ and $\Delta r/r_0$ because
the effect of a larger void can also be obtained by making a smoother, and
hence broader, transition. Current data does not have any sensitivity to
$H_{\rm out}$, because it mainly affects the Hubble rate at very high redshift,
where no good observational data exist. The relative transition width
$\Delta r/r_0$ is also not very well constrained. The only thing we can deduce,
in agreement with the good fit to the data given by the $\Lambda$CDM model,
is that no sudden transition is allowed. The lower limit on $\Delta r/r_0$ is mainly
limited by the Type Ia SN data, and to a lesser extent the BAO.

It should be stressed though, that even though our LTB models give very
good fits with $\chi^2$ that are comparable to that of the $\Lambda$CDM model,
current data {\em do} put significant constraints on the models, and they
will probably be challenged by new observational data in the near
future, and can be falsified.

\begin{table}[h]
\begin{center}
\begin{tabular}{@{}c|c@{ }c@{ }c@{ }c|c|c@{ }c|c@{}}
\hline \hline
Model & $H_0$ & $H_{\rm in}$ & $H_{\rm out}$ & $H_{\rm eff}$
           & $\Omega_{\rm in}$ & $r_0$ & $\Delta r$ & $t_{\rm BB}$ \cr
{\footnotesize units}
           & \multicolumn{4}{c|}{{\footnotesize 100 km s$^{-1}$ Mpc$^{-1}$}} &
           & {\footnotesize Gpc} & $r_0$ & {\footnotesize Gyr} \cr
\hline
GBH&{\footnotesize $-$}&{\footnotesize$0.58\!\pm\!0.03$}
        &{\footnotesize$0.49\!\pm\!0.2$}&{\footnotesize$0.43$}
        &{\footnotesize$0.13\!\pm\!0.06$}&{\footnotesize$2.3\!\pm\!0.9$}
        &{\footnotesize$0.62 (>\!0.20)$}&{\footnotesize14.8}\cr
Constrained&{\footnotesize$0.64\!\pm\!0.03$}&{\footnotesize0.56}
                    &{\footnotesize0.43}&{\footnotesize0.42}
                    &{\footnotesize$0.13\!\pm\!0.06$}&{\footnotesize$2.5\!\pm\!0.7$}
                    &{\footnotesize$0.64 (>\!0.21)$}&{\footnotesize15.3}\cr
\hline \hline
\end{tabular} 
\end{center}
\caption{Best fit values with 2-$\sigma$ error bars for the two models.
The likelihood contours are not closed for $\Delta r$, and only a lower
limit can be given. In the GBH model $H_{\rm out}$ is unconstrained.
For the central values of the other four parameters $H_{\rm out}=0.49$
minimises $\chi^2$. Notice that naturally the best fit GBH model and
the constrained model give similar best fit values, and error bars. Ie.~among
all the different GBH models a model with a homogeneous Big
Bang is preferred.}\label{tab:models}
\end{table}

\begin{figure}
\begin{center}
\includegraphics[width=0.45 \textwidth]{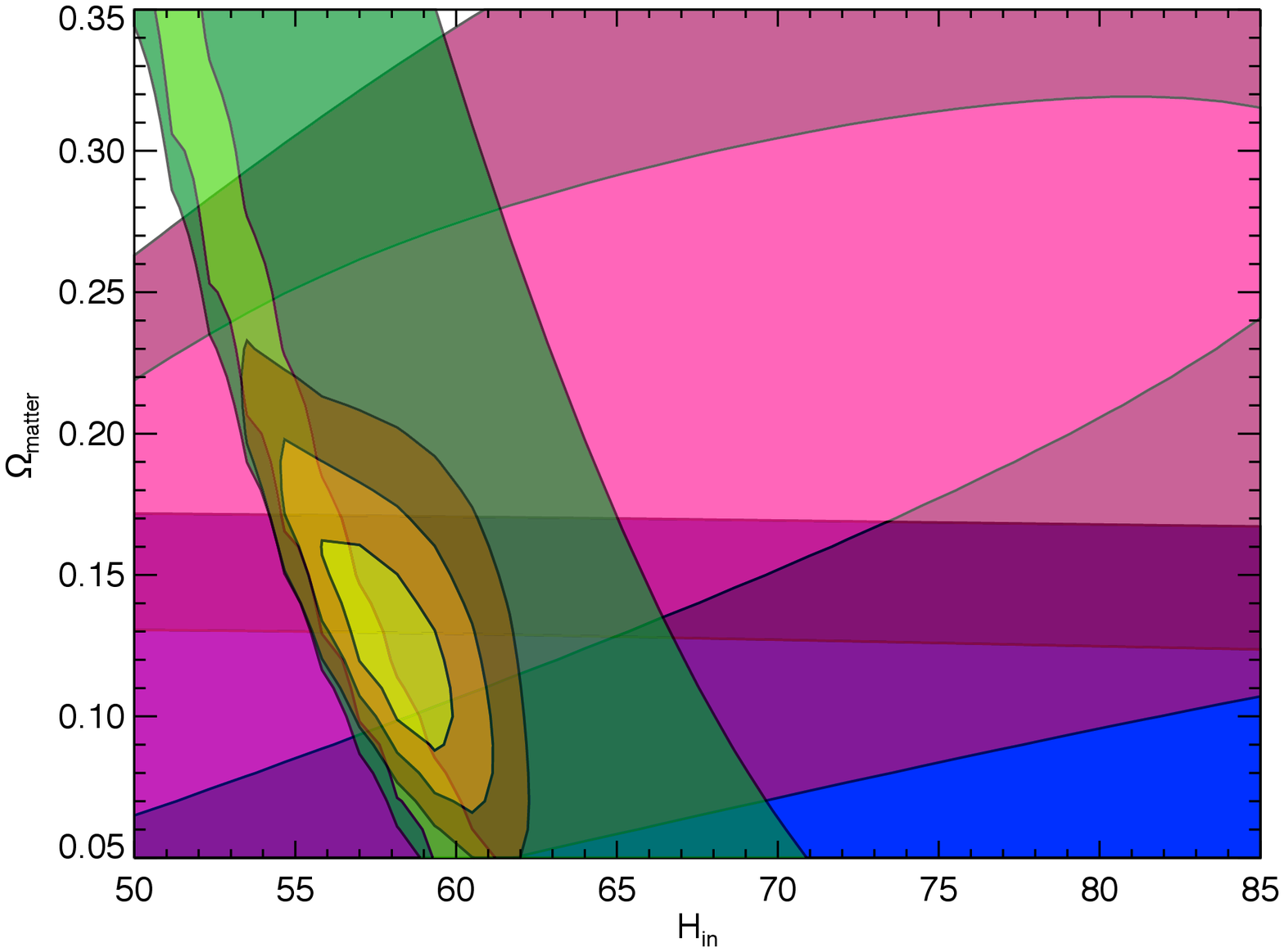}
\includegraphics[width=0.45 \textwidth]{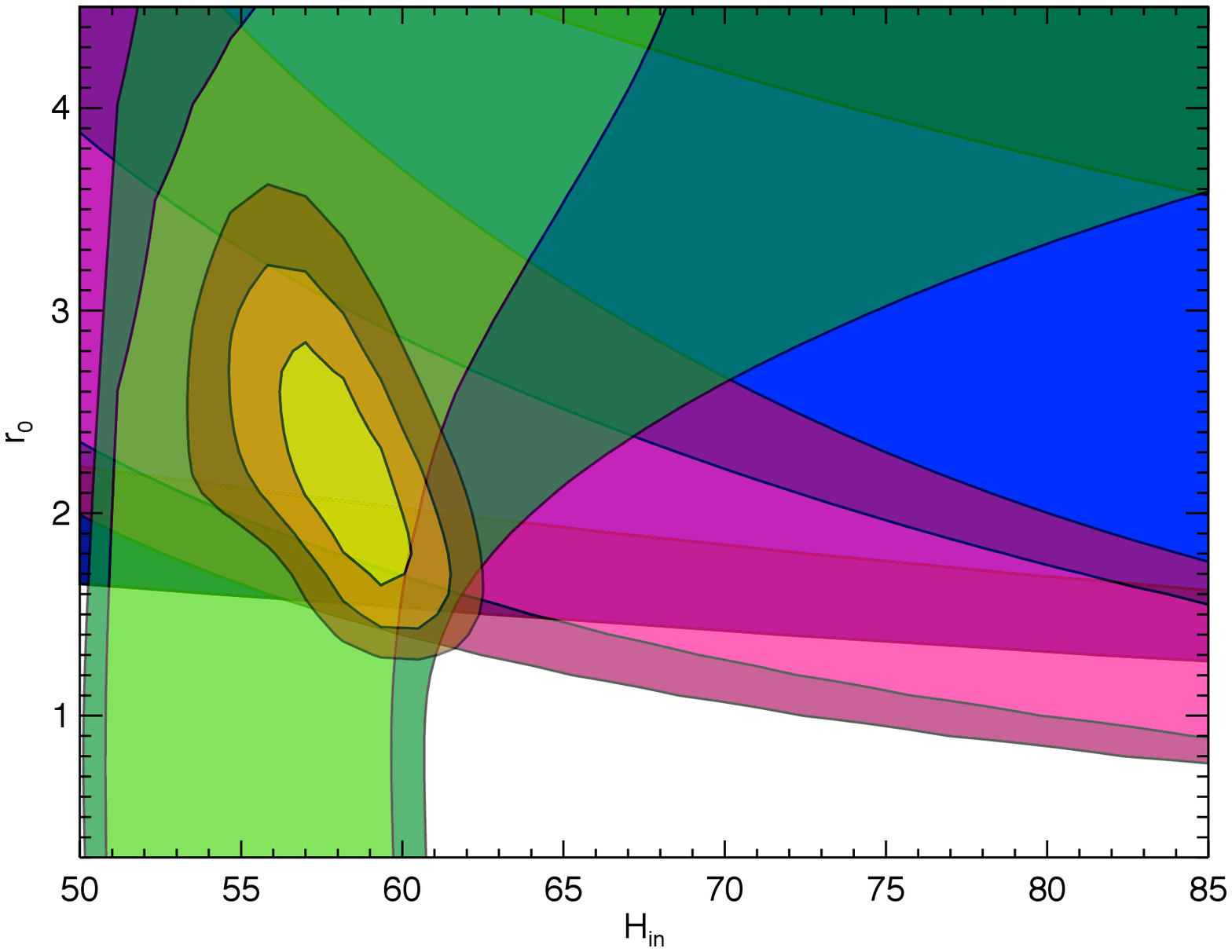}
\includegraphics[width=0.45 \textwidth]{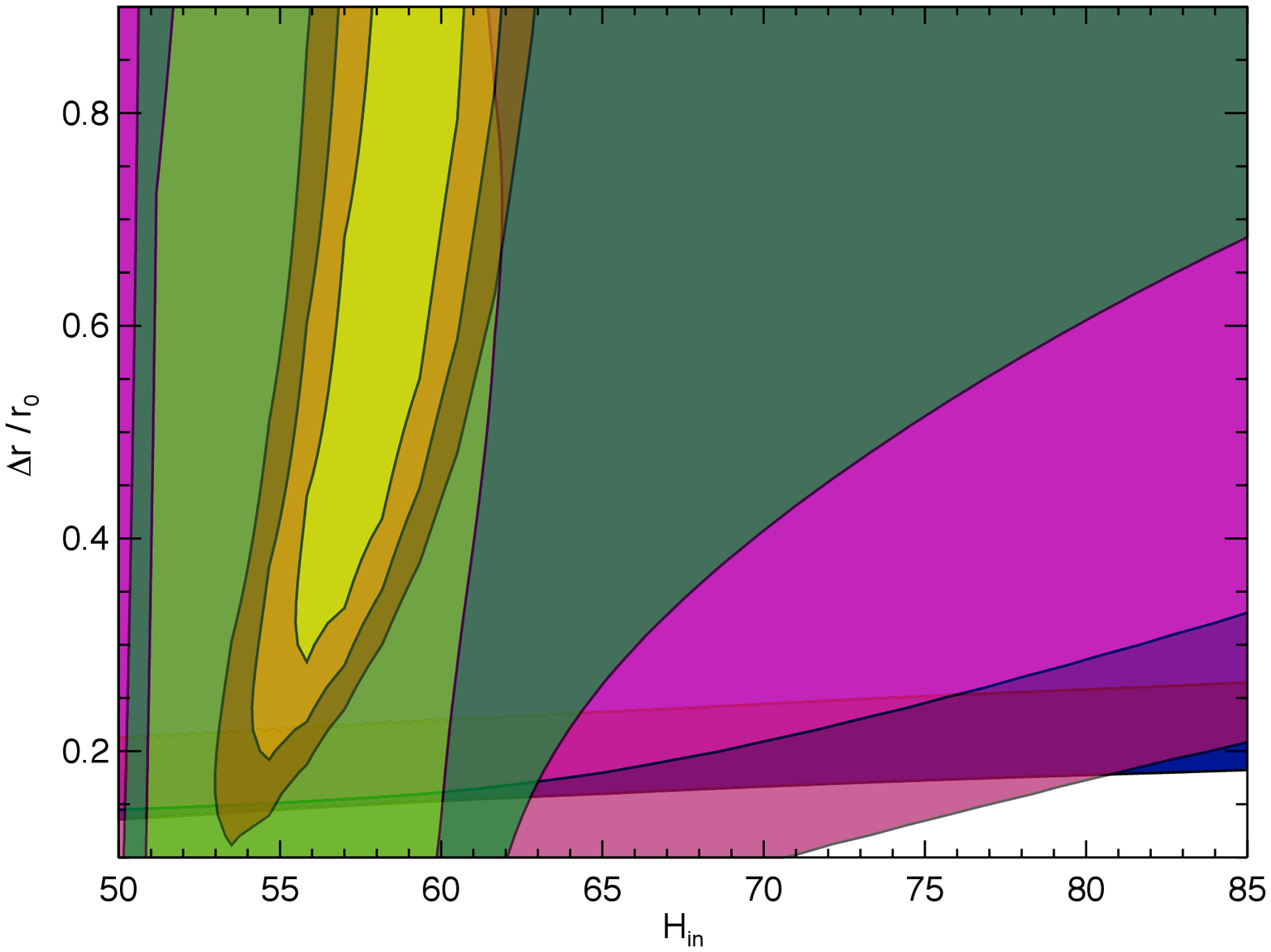}
\includegraphics[width=0.45 \textwidth]{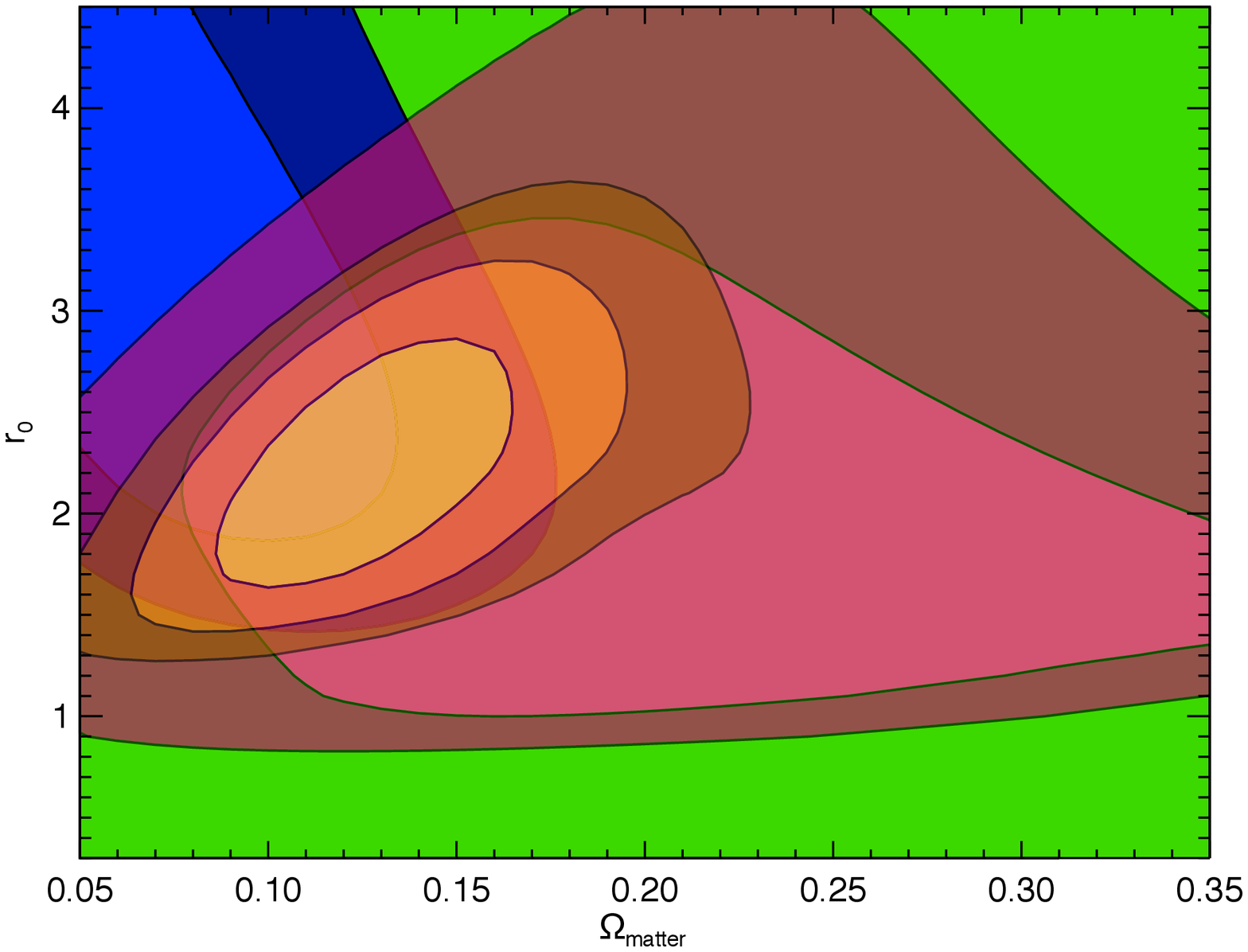}
\includegraphics[width=0.45 \textwidth]{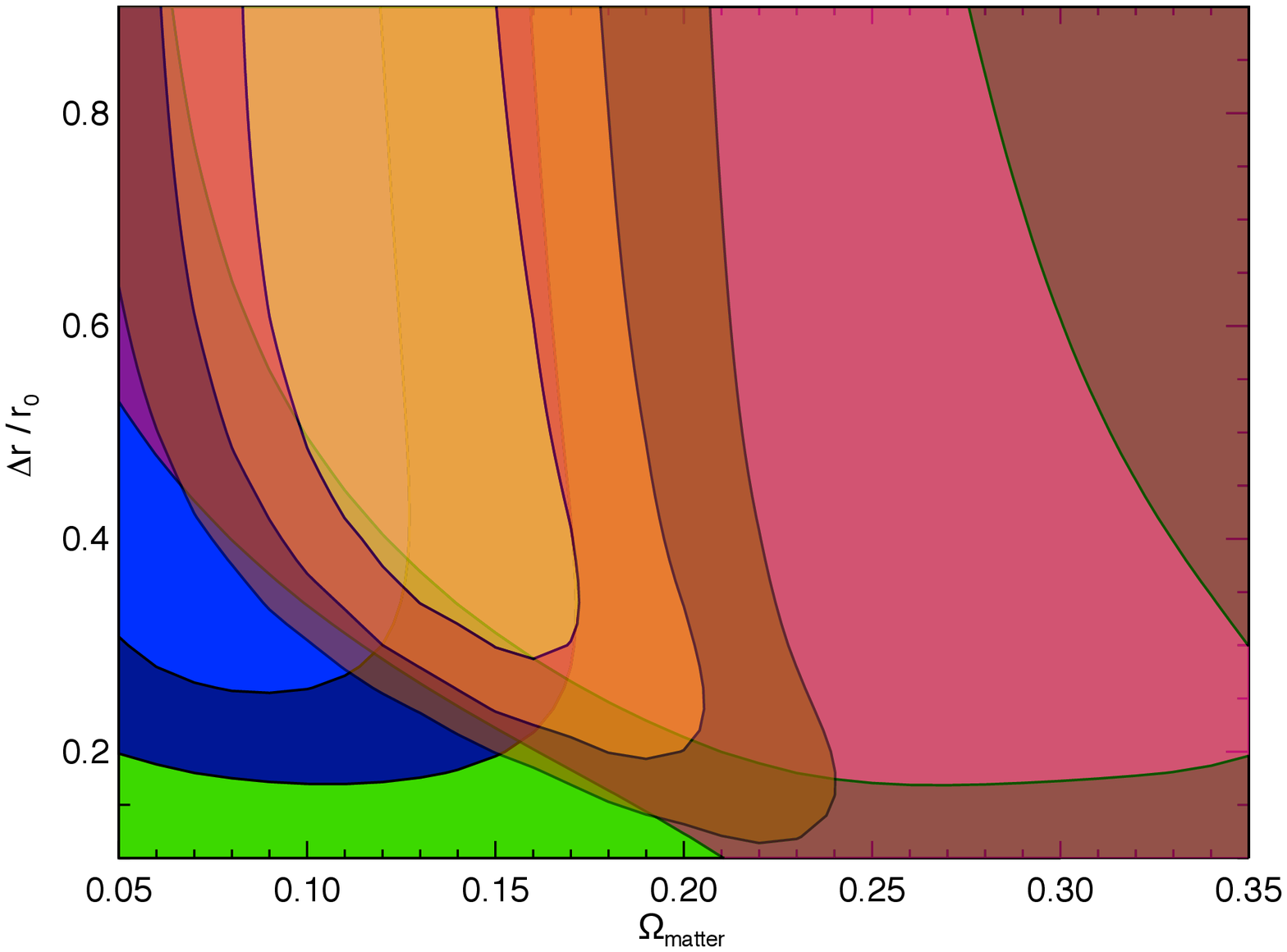}
\includegraphics[width=0.45 \textwidth]{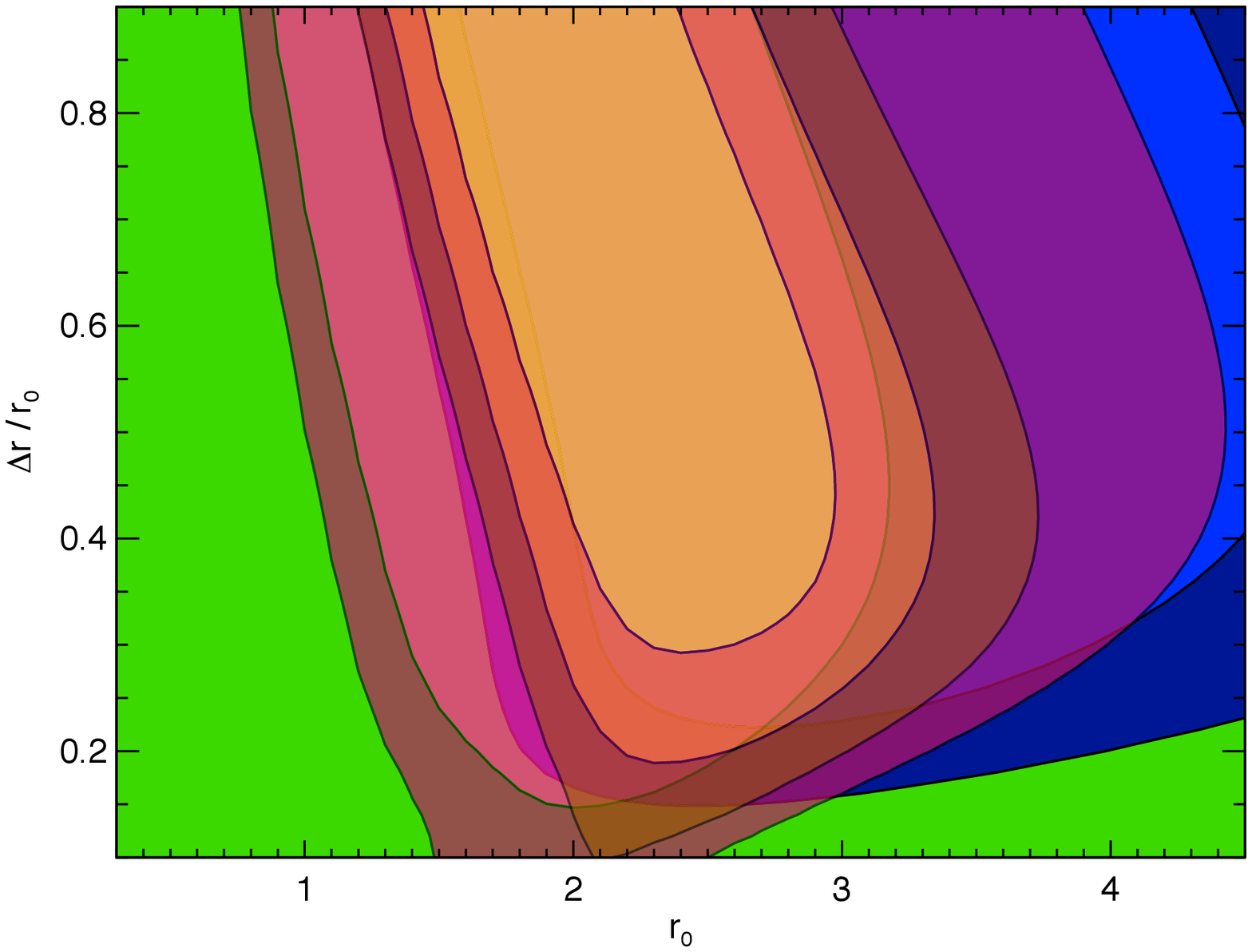}
\caption{\ldots Figure \ref{fig:likelihood2} continued.}
\end{center}
\end{figure}

\begin{figure}
\begin{center}
\includegraphics[width=0.45 \textwidth]{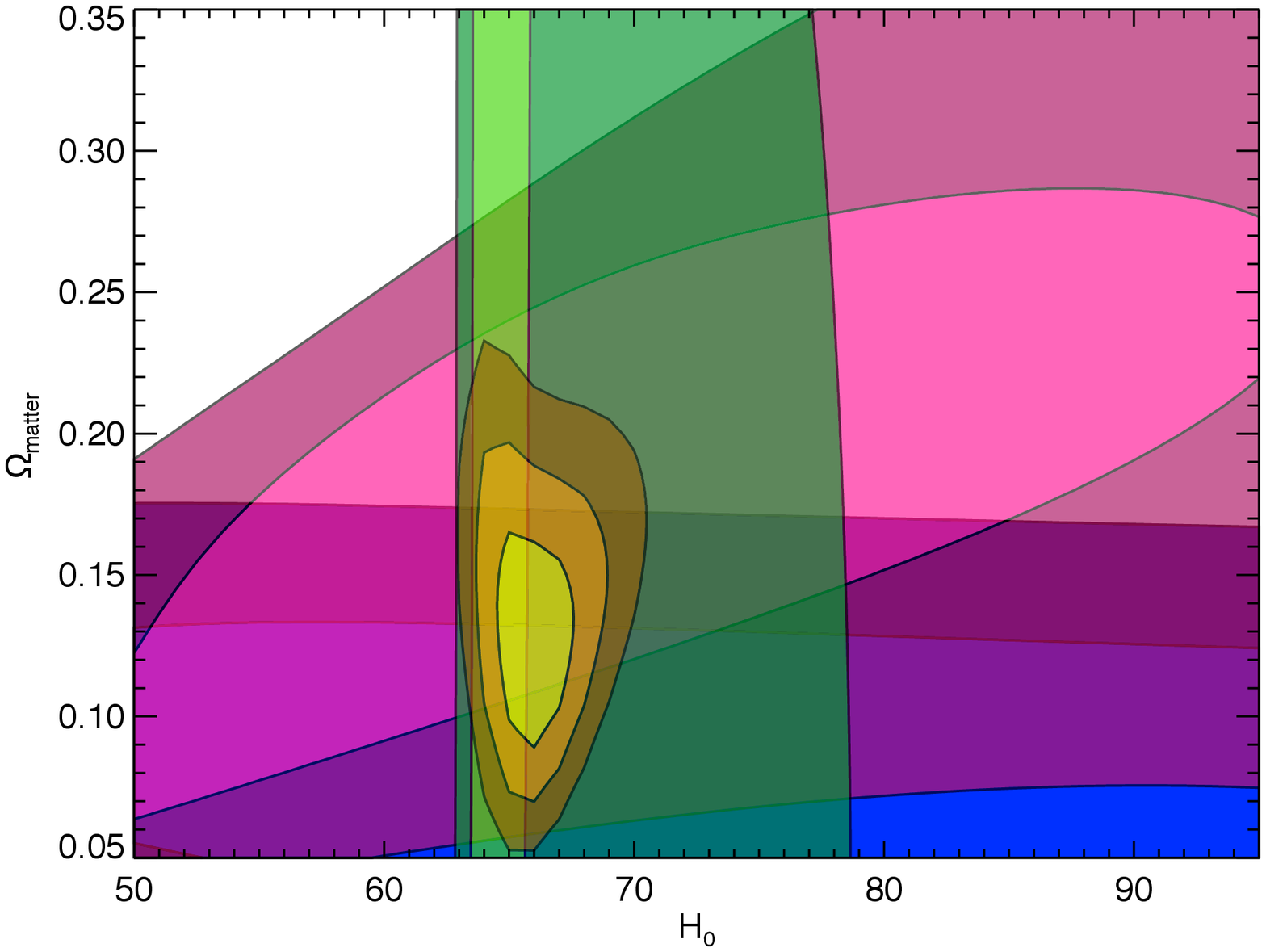}
\includegraphics[width=0.45 \textwidth]{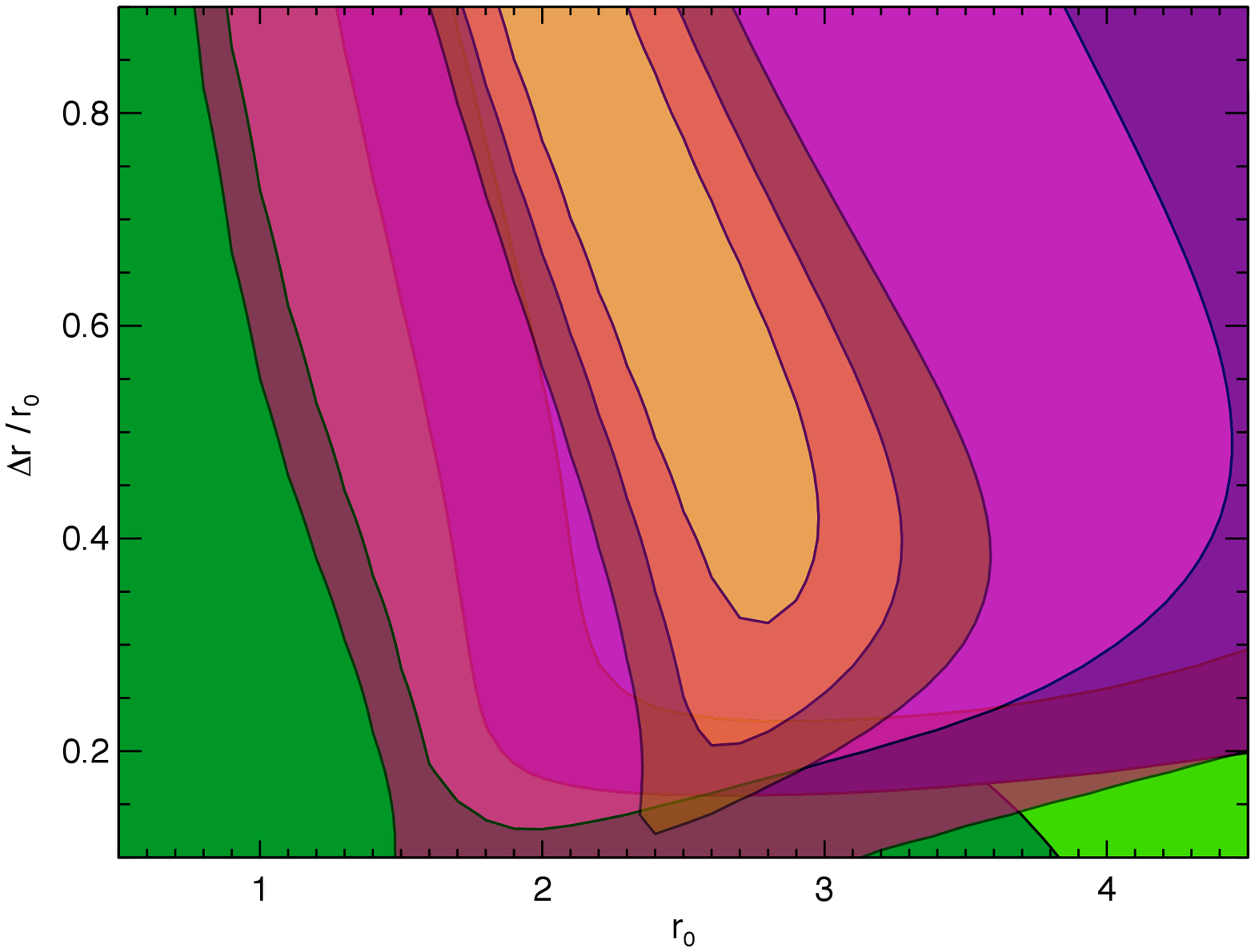}
\includegraphics[width=0.45 \textwidth]{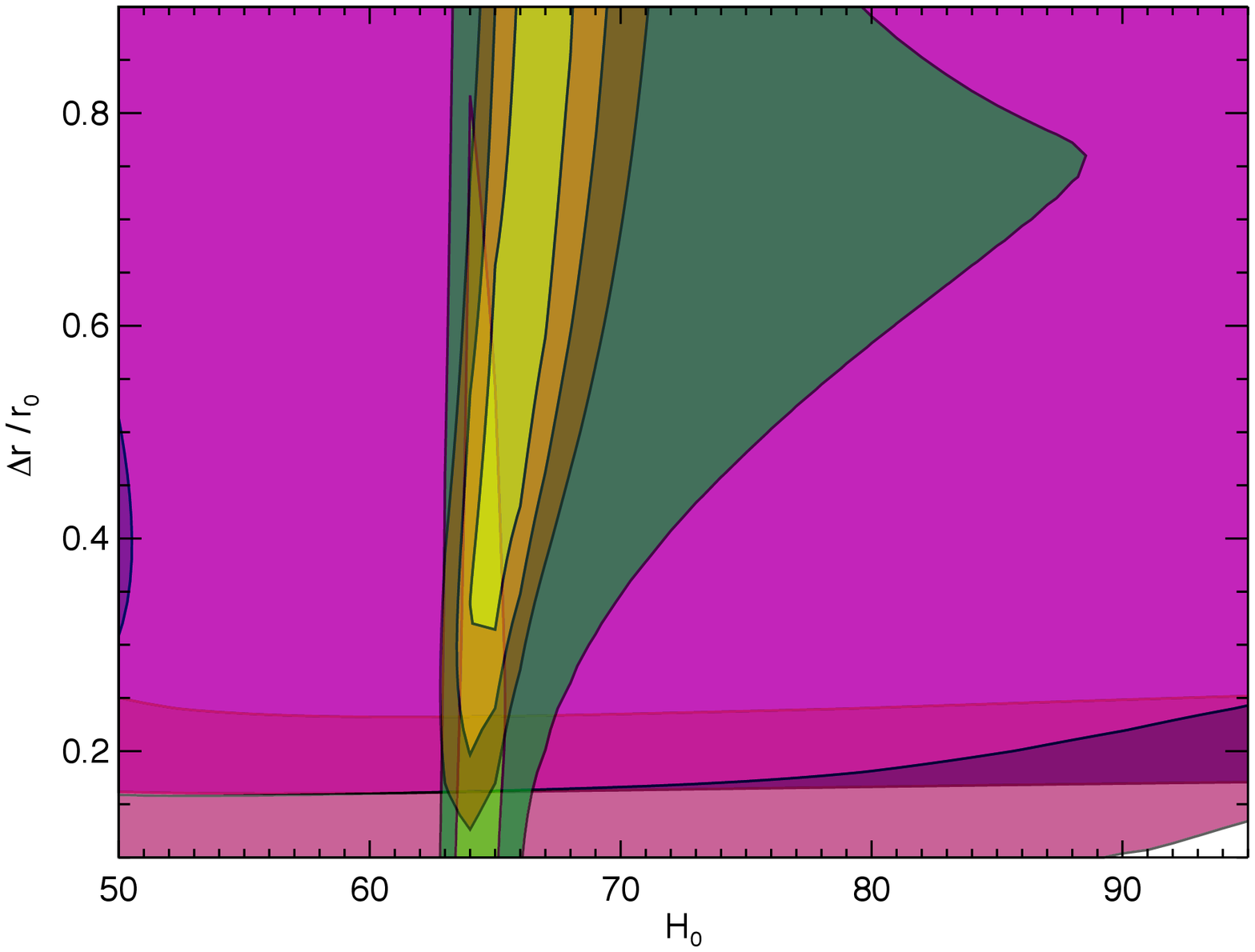}
\includegraphics[width=0.45 \textwidth]{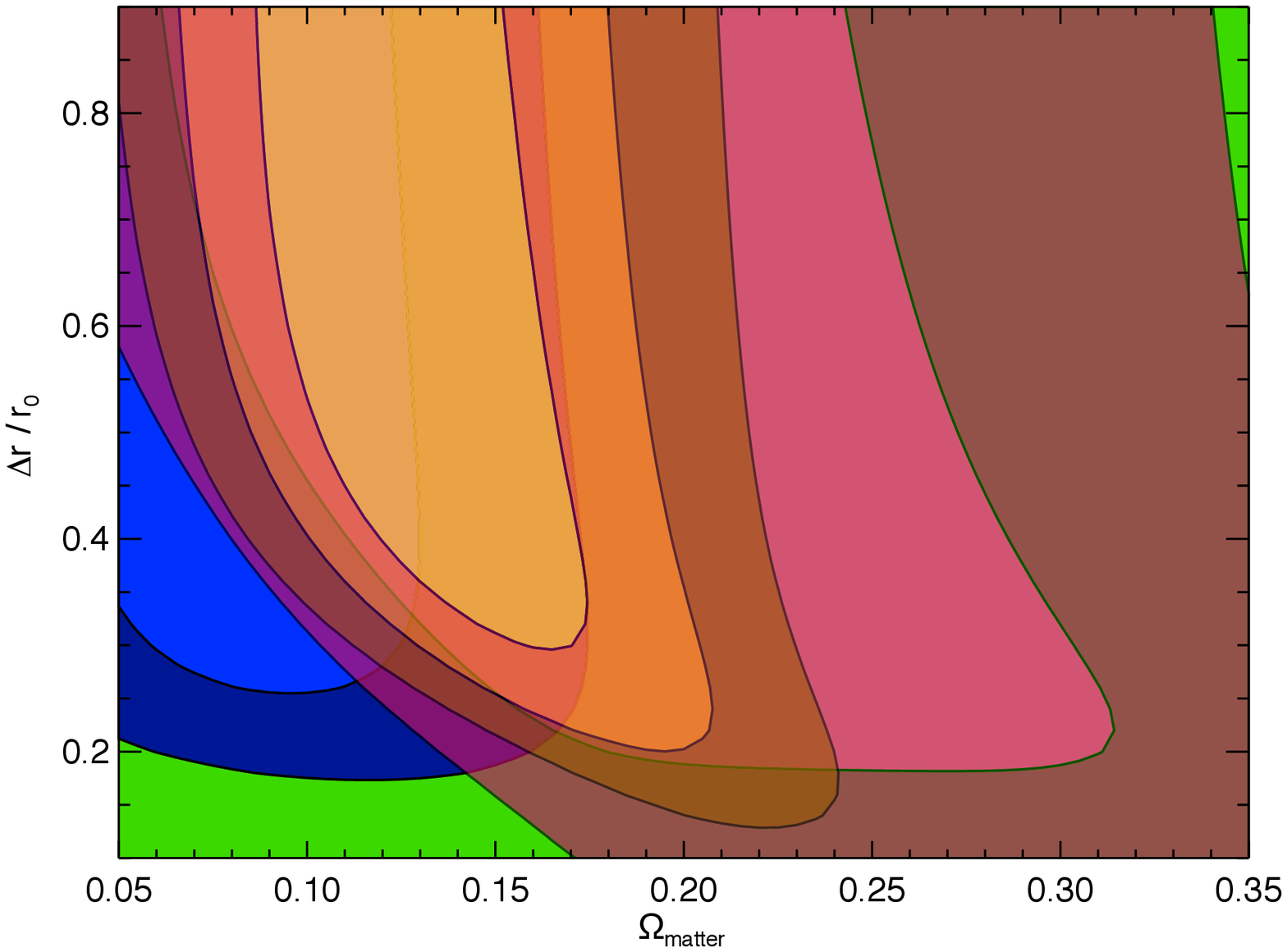}
\includegraphics[width=0.45 \textwidth]{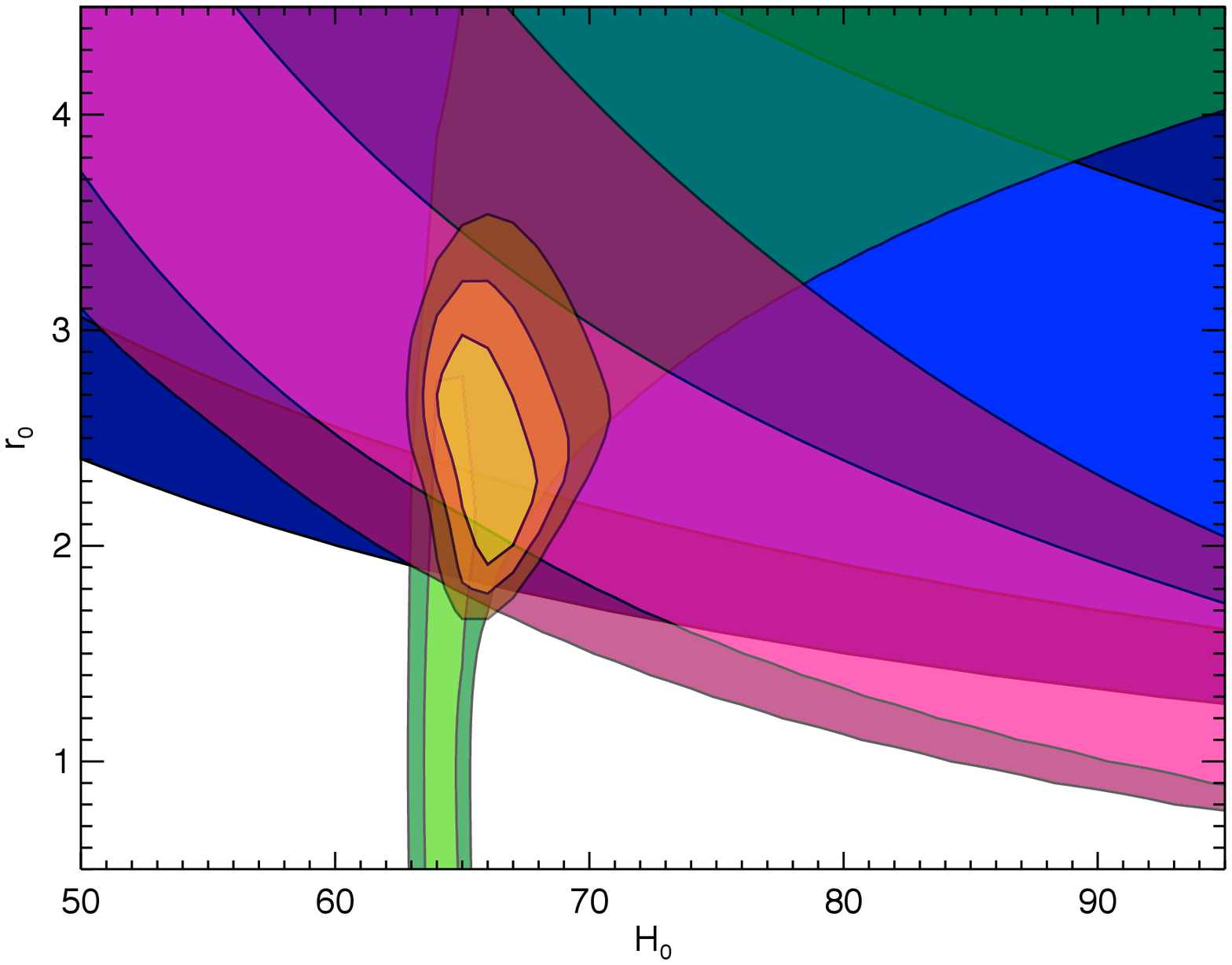}
\includegraphics[width=0.45 \textwidth]{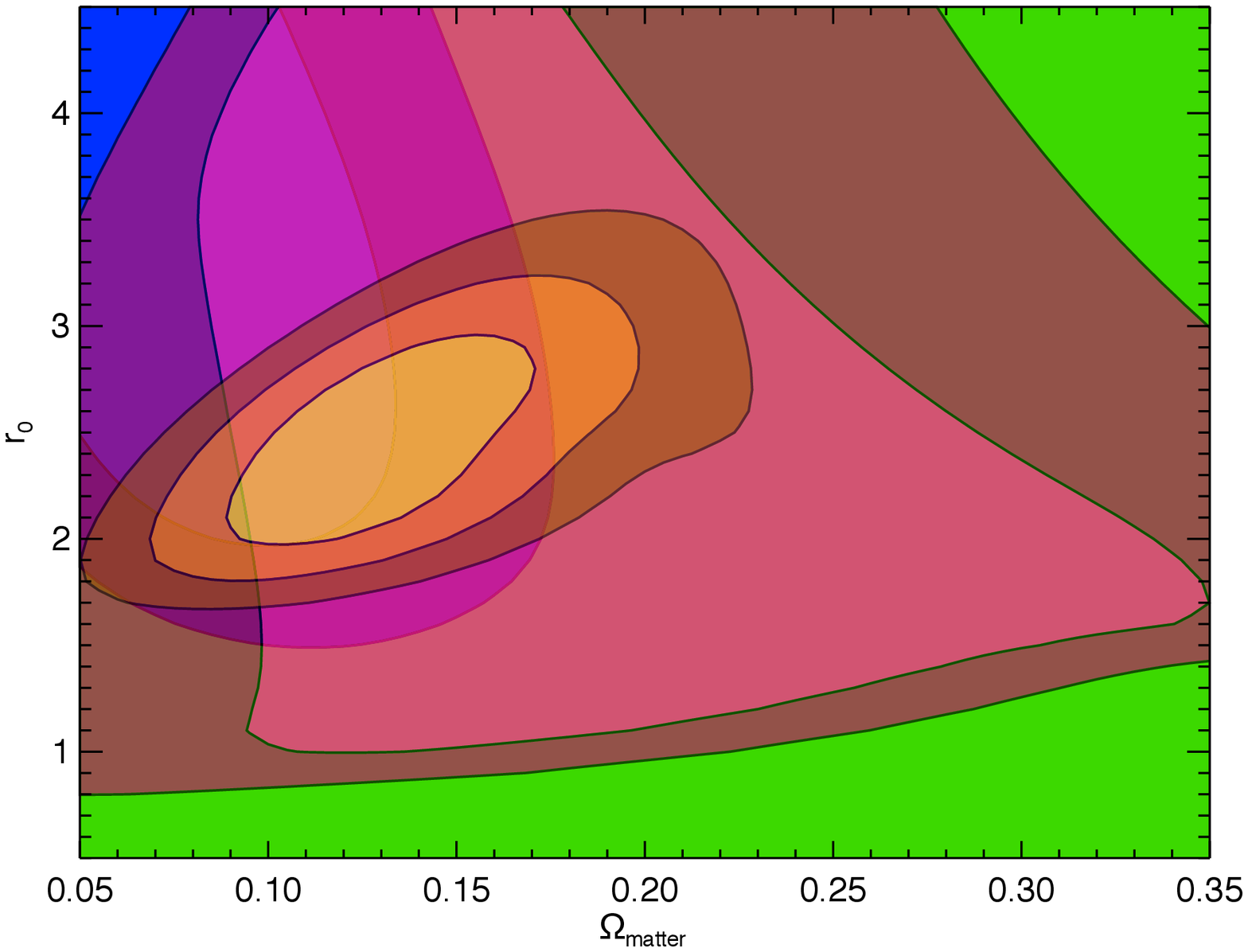}
\caption{Likelihoods for the constrained model: The likelihood for the
combined data set is shown in yellow with 1-, 2-, and 3-$\sigma$
contours, while the individual SNIa, BAO, and CMB data sets are
shown in blue, purple, and green respectively with 1- and
2-$\sigma$ contours.}
\label{fig:likelihood1}
\end{center}
\end{figure}

\subsection{Bayesian analysis}

In this section we would like to find out whether a homogeneous FRW
model of the universe (including the accelerated expansion in terms of
a vacuum energy with constant equation of state $w$) can be used with
confidence when analysing present cosmological data, or should we
rather be more general and assume an inhomogeneous LTB model of the
universe?

The standard frequentist analysis of parameter estimation, given a set
of data, is not very useful for model selection, since it is difficult
to compare models with different number of parameters. For a discussion
about probability theory and model selection see 
Refs.~\cite{Jeffreys,Jaynes,Mackay,D'Agostini}. For instance,
the usual method of comparing minimum $\chi^2$ per effective degree of
freedom normally misses the point and is not very decisive. Other
methods to decide which model gives the best description, given the
data, include various Information Criteria, e.g. Akaike~\cite{Akaike}
and Bayesian~\cite{Schwarz},
which use more or less {\it ad\ hoc} formulae without much
justification and normally do not compare well among eachother.
However, in the last few years there has been a flourishing of several
independent analysis based on the Bayesian evidence associated with a
given likelihood and a given cosmological model, within some given
priors, both theoretical and observational, see e.g. 
Refs.~\cite{Liddle2004,Beltran2005,Parkinson2006,Trotta2007}.

The Bayesian evidence is based on Bayes theorem, which relates the
posterior distribution ${\cal P}(\theta,{\cal M}|{\bf D})$ for the
parameters $\theta$ of the model ${\cal M}$ given the data ${\bf D}$,
in terms of the likelihood distribution function ${\cal L}({\bf D}|
\theta,{\cal M})$ within a given set of priors $\pi(\theta,{\cal M})$,
\begin{equation}\label{BayesThm}
{\cal P}(\theta,{\cal M}|{\bf D}) = {{\cal L}({\bf D}|\theta,{\cal M})
\,\pi(\theta,{\cal M})\over E({\bf D}|{\cal M})}\,,
\end{equation}
where $E$ is the Bayesian evidence, that is the average likelihood
over the priors, 
\begin{equation}\label{BE}
E({\bf D}|{\cal M}) = \int d\theta\ {\cal L}({\bf D}|\theta,{\cal M})\,
\pi(\theta,{\cal M})\,.
\end{equation}
The prior distribution functions contain all the information about the
parameters {\em before} observing the data, e.g. our theoretical
prejudices, our physical intuition about the model, the ranges of
parameters obtained from previous experiments, etc. The Bayesian
evidence is very useful because it allows a comparison of models
within a complete set ${\cal M}_{i=1\dots N}$ (in our case, N=2).
We can compute the posterior probability for each hypothesis (model)
given the data ${\bf D}$ using again Bayes theorem, ${\cal P}({\cal
M}_i| {\bf D})\propto E({\bf D}|{\cal M}_i)\,\pi({\cal M}_i)$, where
$E({\bf D}|{\cal M}_i)$ is the evidence of the data under model ${\cal
M}_i$ and $\pi({\cal M}_i)$ is the prior probability of the $i$-th
model {\em before} we see the data, usually taken to be identical,
i.e. $\pi({\cal M}_{i=1\dots N})=1/N$. Finally, the ratio of the
evidences for two competing models is called the {\em Bayes\ factor},
\begin{equation}\label{BF}
B_{ij} \equiv {E({\bf D}|{\cal M}_i)\over E({\bf D}|{\cal M}_j)}\,.
\end{equation}
This expression provides a mathematical representation of Occam's
razor, because more complex models tend to be less predictive,
lowering their average likelihood (within the priors) in comparison
with simpler, more predictive models. Complex models can only be
favoured if they are able to provide a significantly improved fit to
the data. In simple cases where different models give vastly different
maximum likelihoods there is no need to employ model selection
techniques because they provide only minor corrections to the standard
inference, but they are essential when the difference between maximum
likelihoods is only marginal, as will be the case at hand. The Bayes
factor (\ref{BF}) is then used to give evidence of (i.e.~favour) the
model ${\cal M}_i$ against the model ${\cal M}_j$ using the so called
Jeffreys' scale, a particular interpretation of the Bayes factor which
strengthens its veridic roughly each time the logarithm $\ln B_{ij}$
increases by one unit, from 0 (undecisive) to greater than 5 (strongly
ruled out).

Unfortunately, the computation of the Bayesian evidence (\ref{BE}) is
rather involved and typically requires extensive computational power,
unless the number of parameters is significantly reduced. When the
likelihood of the data given the model parameters is a single isolated
peak, far from the edges of the prior ranges, then there is a simple
approximation to the logarithm of the Bayesian evidence,
\begin{equation}\label{lnE}
\ln E = \ln {\cal L}_{\rm max} - \ln A - \sum_i^n \ln \Delta\theta_i\,, 
\end{equation}
where $A$ is the normalisation of the likelihood, and $\Delta\theta_i
= b_i - a_i$ is the range of parameter $\theta_i \in [a_i,b_i]$,
$i=1\dots n$. Moreover, for the case of a Gaussian likelihood,
$${\cal L}(\theta) = A\,\exp\Big[-{1\over2}{\bf x}^{\rm T}C^{-1} {\bf
x}\Big]\,,$$ we find $A=(2\pi)^{-n/2}/\sqrt{\det C}$, where $C$ is the
covariance matrix and $x_i=\theta_i - \bar\theta_i$. It is clear that
whenever the prior ranges are too big for the likelihood, the Bayesian
evidence is penalised.  Moreover, the more parameters there are, the
larger the difference in $\ln E$, and the larger the logarithm of the
Bayes factor, as expected.

We will now apply this discussion to the case at hand. We will compare
the standard flat $\Lambda$CDM model with a constant equation of state
parameter, $w$, while fixing all other cosmological parameters but
$\Omega_M$. Thus our FRW model has here 2 parameters, which we believe
are the most constraining, since others like the local rate of
expansion do not provide significant extra information. We performed a
calculation of the minimum $\chi^2$ within a grid of 5000 models,
where $w\in[-2.0, 0.0]$ and $\Omega_M\in[0.0,0.6]$. The maximum
likelihood corresponds to a $\chi^2_{\rm min}(\Lambda CDM) = 197.05$
for $w = - 1.005$ and $\Omega_M = 0.276$.

On the other hand, the constrained GBH model has 4 independent
parameters, the local rate of expansion $H_0 = 100\,h$ km/s/Mpc, the
local matter fraction $\Omega_m\equiv\Omega_M(0)$, the transition
distance $r_0$, and the width $\Delta r/r_0$, providing a grid of
several million models within the ranges $h\in[0.50,0.95]$,
$\Omega_m\in[0.05,0.35]$, $r_0\in[0.5,4.5]$ and $\Delta r/r_0 \in
[0.1,0.9]$. We found that the maximum likelihood corresponds to a
$\chi^2_{\rm min}(GBH) = 197.845$ for $h=0.659$, $\Omega_m = 0.124$,
$r_0 = 2.47$ Gpc and $\Delta r/r_0 = 0.638$.

At face value it seems that the inhomogeneous model provides as good a fit
to the data as the FRW one. If we compute the usual minimum $\chi^2$ per
effective number of degrees of freedom (i.e.~the number of data points
minus the number of parameters), we find
\begin{equation}
\chi^2_{\rm min}(\Lambda CDM)/d.o.f. = 1.021\,, \hspace{1cm}
\chi^2_{\rm min}(GBH)/d.o.f. = 1.036\,,
\end{equation}
so that both models seem excellent descriptions of the data, the
first one being slightly better. However, there are other indicators
more appropriate for model comparison, like the (corrected) 
Akaike Information Criterion ($AIC$), computed as
\begin{equation}
AIC = \chi^2_{\rm min} + 2k + {2k(k-1)\over N-k-1}\,,
\end{equation}
where $k$ is the number of parameters and $N$ is the number of data
points.  In our case this gives $AIC(\Lambda CDM) = 201.1$, while
$AIC(GBH) = 206.0$, which would clearly favour the homogeneous
FRW model. On the other hand, if we choose to compare models with the
Bayesian Information Criterion ($BIC$), computed as
\begin{equation}
BIC = \chi^2_{\rm min} + k\ln N\,,
\end{equation}
we find $BIC(\Lambda CDM) = 207.6$, while $BIC(GBH) = 218.9$,
which would very strongly favour the homogeneous FRW model. Clearly,
neither method gives a good assessment for choosing among models.
This is the reason why the Bayesian evidence has been used recently in
the context of model comparison.

If we compute the Bayes factor (\ref{BF}) by performing the integral
of the likelihood over the priors, (\ref{BE}), we find
\begin{equation}
\ln E(\Lambda CDM) = -103.1\,, \hspace{1.5cm}
\ln E(GBH) = -106.7\,,
\end{equation}
and therefore the logarithm of the Bayes factor is $\ln B_{12} = 3.6$,
which clearly favours the homogeneous FRW model against the GBH-LTB
model. It seems that the bayesian evidence method discards
significantly, but not very strongly, the inhomogeneous model versus
the usual FRW model. It is possible that, in the future, better data
sets and stronger priors on cosmological parameters may discard once
and for all the inhomogeneous model. For the moment, a local void with
a size of several Gpc, with matter well below average, and a local
rate of expansion of 71 km/s/Mpc, can account for both the distant
(CMB), intermediate (SNIa) and local (BAO) data sets.

\section{Discussion and Conclusions}

We have shown that present observations do not exclude the possibility
that we live close to the center of a large void. This is a appealing
possibility which effectively gets rid of the necessity to introduce
an ad hoc cosmological constant in our model of the universe. Moreover,
it is consistent with early universe cosmology in terms of inflationary
initial conditions for the origin of large scale structures. Perhaps these
voids arise due to large non-perturbative inhomogeneities associated  
with the stochastic nature of the inflaton evolution~\cite{Linde:1994gy}, or due
to large non-gaussianities in the primordial spectrum coming from inflation
that could arise due to phase transitions or in multifield inflation
\cite{Kofman:1986wm,Chen:2008wn}. 
In fact, we already know there can be other voids in our local patch of
the universe, with a Gpc scale, as exemplified by the observed cold
spot in the CMB~\cite{Cruz:2006fy}.

We have analysed the likelihood of such an interpretation of the
present acceleration of the universe, using data from the Cosmic 
Microwave Background (on large scales), Supernova Ia (at intermediate 
scales), Baryon Acoustic Oscillations, the present age and the local 
rate of expansion (at small scales). All the data seems to be
consistent at the 95\% confidence level with a local void of size
around 2.5 Gpc, within an Einstein-de Sitter universe on large scales,
without the need to introduce a cosmological constant. The apparent
acceleration of the universe can be interpreted here as due to the 
curved path of photons in this locally open universe.
We should however take into account that a possible displacement from
the centre of the void will induce a dipole in the CMB \cite{Piran,Alnes:2006pf}.
If this (geometrically induced) dipole is not cancelled by a peculiar velocity
towards the centre, to be in agreement with the opbserved CMB dipole, one can show
\cite{Alnes:2006pf} that we cannot be displaced more than approximately
100 Mpc from the centre. From a probabilistic point of view, this is a generic
argument against Gpc sized voids.

We have performed a Bayesian analysis in order to compare two 
competing models, the predominant $\Lambda$CDM model and our GBH
inhomogeneous model. While the usual frequentists analysis does
not discard the GBH model against the $\Lambda$CDM model, there 
seems to be strong but not decisive (bayesian) evidence against the 
GBH model. It is possible that in the near future, with much better
cosmological data on large scale structures and an extended set of
supernovae at intermediate and high redshift, we may be able to constrain and
definitely rule out the inhomogeneous LTB model. We should also 
mention that the data seems to favour a homogeneous Big Bang since 
the constrained GBH model gives the same likelihood contours and 
minimum $\chi^2$ than the unconstrained model, while having one
parameter less.

At the moment we are studying the effect that a generic LTB model 
has on the growth of structure in order to constrain further the
GBH model with data from ISW-LSS correlations and the Lyman-$\alpha$
forest within the Alcock-Paczynski-test analysis~\cite{Alicia}.

In conclusion, we cannot discard that we live in an inhomogeneous 
local void within an asymptotically Einstein-de Sitter universe.
The possibility that we have misinterpreted the present acceleration
and that the cosmological constant is nothing but a 
mirage has been addressed 
recently~\cite{Rasanen:2003,Mattsson:2007tj,Wiltshire:2007zj,Li:2008yj}.
We have added to the discussion the comparison with a large, albeit 
incomplete, set of cosmological observations, and the bayesian
analysis appropriate for model selection. We hope in the future
to provide further constraints on the model and possibly rule it out.

\section*{Acknowledgements}

We thank Andrei Linde, Andy Albrecht, David Valls-Gabaud, Sysky R\"as\"anen, and
Alex Kusenko for generous comments and suggestions. We also thank the Danish
Centre of Scientific Computing (DCSC) for granting the computer resources used.
JGB would like to thank the Kavli Institute for Theoretical Physics
for hospitality during the last stages of the work, supported 
in part by the National Science Foundation under Grant No. PHY05-51164.
We also acknowledge financial support from the Spanish Research
Ministry (M.E.C.), under the contract FPA2006-05807.

\section*{References}
\bibliographystyle{hunsrt}
\bibliography{paper}

\begin{thebibliography}{10}

\bibitem{des}
\url{http://www.darkenergysurvey.org}.

\bibitem{landau}
\url{http://www.aip.org/web_bin/pt/vol-58/iss-8/p53.shtml}.

\bibitem{Lemaitre:1997ab}
George Lema\^itre.
\newblock {"The Expanding Universe"}.
\newblock {\em Gen. Rel. Grav.}, 29:641--680, 1997.

\bibitem{Tolman:1934za}
Richard~C. Tolman.
\newblock {"Effect of imhomogeneity on cosmological models"}.
\newblock {\em Proc. Nat. Acad. Sci.}, 20:169--176, 1934.

\bibitem{Bondi:1947av}
H.~Bondi.
\newblock {"Spherically symmetrical models in general relativity"}.
\newblock {\em Mon. Not. Roy. Astron. Soc.}, 107:410--425, 1947.

\bibitem{Zehavi:1998gz}
Idit Zehavi, Adam~G. Riess, Robert~P. Kirshner, and Avishai Dekel.
\newblock {"A Local Hubble Bubble from SNe Ia?"}.
\newblock {\em Astrophys. J.}, 503:483, 1998, astro-ph/9802252.

\bibitem{Tomita:2000rf}
Kenji Tomita.
\newblock {"Anisotropy of the Hubble Constant in a Cosmological Model with a
  Local Void on Scales of ~ 200 Mpc"}.
\newblock 2000, astro-ph/0005031.

\bibitem{Tomita:2000jj}
Kenji Tomita.
\newblock {"A Local Void and the Accelerating Universe"}.
\newblock {\em Mon. Not. Roy. Astron. Soc.}, 326:287, 2001, astro-ph/0011484.

\bibitem{Tomita:2001gh}
Kenji Tomita.
\newblock {"Analyses of Type Ia Supernova Data in Cosmological Models with a
  Local Void"}.
\newblock {\em Prog. Theor. Phys.}, 106:929--939, 2001, astro-ph/0104141.

\bibitem{Frith:2003tb}
William~J. Frith, G.~S. Busswell, R.~Fong, N.~Metcalfe, and T.~Shanks.
\newblock {"The Local Hole in the Galaxy Distribution: Evidence from 2MASS"}.
\newblock {\em Mon. Not. Roy. Astron. Soc.}, 345:1049, 2003, astro-ph/0302331.

\bibitem{Busswell:2003ta}
G.~S. Busswell et~al.
\newblock {"The Local Hole in the Galaxy Distribution: New Optical Evidence"}.
\newblock {\em Mon. Not. Roy. Astron. Soc.}, 354:991, 2004, astro-ph/0302330.

\bibitem{Vielva:2003et}
Patricio Vielva, E.~Martinez-Gonzalez, R.~B. Barreiro, J.~L. Sanz, and
  L.~Cayon.
\newblock {"Detection of non-Gaussianity in the WMAP 1-year data using
  spherical wavelets"}.
\newblock {\em Astrophys. J.}, 609:22--34, 2004, astro-ph/0310273.

\bibitem{Cruz:2006sv}
M.~Cruz, M.~Tucci, E.~Martinez-Gonzalez, and P.~Vielva.
\newblock {"The non-Gaussian Cold Spot in WMAP: significance, morphology and
  foreground contribution"}.
\newblock {\em Mon. Not. Roy. Astron. Soc.}, 369:57--67, 2006,
  astro-ph/0601427.

\bibitem{Cruz:2006fy}
Marcos Cruz, L.~Cayon, E.~Martinez-Gonzalez, P.~Vielva, and J.~Jin.
\newblock {"The non-Gaussian Cold Spot in the 3-year WMAP data"}.
\newblock {\em Astrophys. J.}, 655:11--20, 2007, astro-ph/0603859.

\bibitem{Linde:1994gy}
Andrei~D. Linde, Dmitri~A. Linde, and Arthur Mezhlumian.
\newblock {"Do we live in the center of the world?"}.
\newblock {\em Phys. Lett.}, B345:203--210, 1995, hep-th/9411111.

\bibitem{Linde:1993xx}
Andrei~D. Linde, Dmitri~A. Linde, and Arthur Mezhlumian.
\newblock {"From the Big Bang theory to the theory of a stationary universe"}.
\newblock {\em Phys. Rev.}, D49:1783--1826, 1994, gr-qc/9306035.

\bibitem{GarciaBellido:1993wn}
Juan Garcia-Bellido, Andrei~D. Linde, and Dmitri~A. Linde.
\newblock {"Fluctuations of the gravitational constant in the inflationary
  Brans-Dicke cosmology"}.
\newblock {\em Phys. Rev.}, D50:730--750, 1994, astro-ph/9312039.

\bibitem{Enqvist:2006cg}
Kari Enqvist and Teppo Mattsson.
\newblock {"The effect of inhomogeneous expansion on the supernova
  observations"}.
\newblock {\em JCAP}, 0702:019, 2007, astro-ph/0609120.

\bibitem{Enqvist:2007vb}
Kari Enqvist.
\newblock {"Lemaitre-Tolman-Bondi model and accelerating expansion"}.
\newblock 2007, arXiv:0709.2044 [astro-ph].

\bibitem{Alnes:2005rw}
Havard Alnes, Morad Amarzguioui, and Oyvind Gron.
\newblock {"An inhomogeneous alternative to dark energy?"}.
\newblock {\em Phys. Rev.}, D73:083519, 2006, astro-ph/0512006.

\bibitem{Hellaby:2005}
Charles Hellaby and Andrzej Krasinski.
\newblock {Alternative methods of describing structure formation in the
  Lemaitre-Tolman model}.
\newblock {\em Phys. Rev.}, D73:023518, 2006, gr-qc/0510093.

\bibitem{Linder:2003dr}
Eric~V. Linder and Adrian Jenkins.
\newblock {"Cosmic Structure and Dark Energy"}.
\newblock {\em Mon. Not. Roy. Astron. Soc.}, 346:573, 2003, astro-ph/0305286.

\bibitem{Linder:2005in}
Eric~V. Linder.
\newblock {"Cosmic growth history and expansion history"}.
\newblock {\em Phys. Rev.}, D72:043529, 2005, astro-ph/0507263.

\bibitem{SDSS-SN}
\url{http://sdssdp47.fnal.gov/sdsssn/sdsssn.html}.

\bibitem{PAU}
\url{http://www.ice.csic.es/pau}.

\bibitem{WMAP3}
D.~N. Spergel et~al.
\newblock {"Wilkinson Microwave Anisotropy Probe (WMAP) three year results:
  Implications for cosmology"}.
\newblock {\em Astrophys. J. Suppl.}, 170:377, 2007, astro-ph/0603449.

\bibitem{Eisenstein:1997ik}
Daniel~J. Eisenstein and Wayne Hu.
\newblock {"Baryonic Features in the Matter Transfer Function"}.
\newblock {\em Astrophys. J.}, 496:605, 1998, astro-ph/9709112.

\bibitem{Eisenstein:2005su}
Daniel~J. Eisenstein et~al.
\newblock {"Detection of the Baryon Acoustic Peak in the Large-Scale
  Correlation Function of SDSS Luminous Red Galaxies"}.
\newblock {\em Astrophys. J.}, 633:560--574, 2005, astro-ph/0501171.

\bibitem{Okumura:2007br}
Teppei Okumura et~al.
\newblock {"Large-Scale Anisotropic Correlation Function of SDSS Luminous Red
  Galaxies"}.
\newblock 2007, arXiv:0711.3640 [astro-ph].

\bibitem{Percival:2007yw}
Will~J. Percival et~al.
\newblock {"Measuring the Baryon Acoustic Oscillation scale using the SDSS and
  2dFGRS"}.
\newblock 2007, arXiv:0705.3323 [astro-ph].

\bibitem{Padmanabhan:2006ku}
Nikhil Padmanabhan et~al.
\newblock {"The Clustering of Luminous Red Galaxies in the Sloan Digital Sky
  Survey Imaging Data"}.
\newblock {\em Mon. Not. Roy. Astron. Soc.}, 378:852--872, 2007,
  astro-ph/0605302.

\bibitem{Alicia}
Alicia Bueno, Juan Garc\'{\i}a-Bellido, and Troels Haugboelle.
\newblock {"Linear perturbation theory in LTB models"}.
\newblock in preparation.

\bibitem{Davis:2007na}
Tamara~M. Davis et~al.
\newblock {"Scrutinizing exotic cosmological models using {ESSENCE} supernova
  data combined with other cosmological probes"}.
\newblock {\em Astrophys. J.}, 666:716, 2007, astro-ph/0701510.

\bibitem{Jha:2006fm}
Saurabh Jha, Adam~G. Riess, and Robert~P. Kirshner.
\newblock {"Improved Distances to Type Ia Supernovae with Multicolor Light
  Curve Shapes: MLCS2k2"}.
\newblock {\em Astrophys. J.}, 659:122--148, 2007, astro-ph/0612666.

\bibitem{Astier:2005qq}
Pierre Astier et~al.
\newblock {"The Supernova Legacy Survey: Measurement of $\Omega_M$,
  $\Omega_\Lambda$ and $w$ from the First Year Data Set"}.
\newblock {\em Astron. Astrophys.}, 447:31--48, 2006, astro-ph/0510447.

\bibitem{WoodVasey:2007jb}
W.~Michael Wood-Vasey et~al.
\newblock {"Observational Constraints on the Nature of the Dark Energy: First
  Cosmological Results from the ESSENCE Supernova Survey"}.
\newblock {\em Astrophys. J.}, 666:694, 2007, astro-ph/0701041.

\bibitem{Riess:2006fw}
Adam~G. Riess et~al.
\newblock {"New Hubble Space Telescope Discoveries of Type Ia Supernovae at $z
  > 1$: Narrowing Constraints on the Early Behavior of Dark Energy"}.
\newblock {\em Astrophys. J.}, 659:98, 2006, astro-ph/0611572.

\bibitem{Krauss:2003em}
Lawrence~M. Krauss and Brian Chaboyer.
\newblock {"Age Estimates of Globular Clusters in the Milky Way: Constraints on
  Cosmology"}.
\newblock {\em Science}, 299:65--70, 2003.

\bibitem{Freedman:2000cf}
W.~L. Freedman et~al.
\newblock {"Final Results from the Hubble Space Telescope Key Project to
  Measure the Hubble Constant"}.
\newblock {\em Astrophys. J.}, 553:47--72, 2001, astro-ph/0012376.

\bibitem{Allen:2007ue}
S.~W. Allen et~al.
\newblock {"Improved constraints on dark energy from Chandra X-ray observations
  of the largest relaxed galaxy clusters"}.
\newblock 2007, arXiv:0706.0033 [astro-ph].

\bibitem{Celerier:1999hp}
Marie-Noelle Celerier.
\newblock {"Do we really see a cosmological constant in the supernovae data?"}.
\newblock {\em Astron. Astrophys.}, 353:63--71, 2000, astro-ph/9907206.

\bibitem{Hinshaw:2006ia}
G.~Hinshaw et~al.
\newblock {"Three-year Wilkinson Microwave Anisotropy Probe (WMAP)
  observations: Temperature analysis"}.
\newblock {\em Astrophys. J. Suppl.}, 170:288, 2007, astro-ph/0603451.

\bibitem{Hunt:2007dn}
Paul Hunt and Subir Sarkar.
\newblock {"Multiple inflation and the WMAP 'glitches' II. Data analysis and
  cosmological parameter extraction"}.
\newblock 2007, arXiv:0706.2443 [astro-ph].

\bibitem{Alexander:2007xx}
Stephon Alexander, Tirthabir Biswas, Alessio Notari, and Deepak Vaid.
\newblock {Local Void vs Dark Energy: Confrontation with WMAP and Type Ia
  Supernovae}.
\newblock 2007, arXiv:0712.0370 [astro-ph].

\bibitem{Page:2003}
L.~Page et~al.
\newblock {First Year Wilkinson Microwave Anisotropy Probe (WMAP) Observations:
  Interpretation of the TT and TE Angular Power Spectrum Peaks}.
\newblock {\em Astrophys. J. Suppl.}, 148:233, 2003, astro-ph/0302220.

\bibitem{Jeffreys}
H.~Jeffreys.
\newblock {"The theory of probability"}.
\newblock Oxford U.P. (1998).

\bibitem{Jaynes}
E.T. Jaynes.
\newblock {"Probability Theory: the Logic of Science"}.
\newblock Cambridge U.P. (2003).

\bibitem{Mackay}
D.J.C. Mackay.
\newblock {"Information theory, inference and learning algorithms"}.
\newblock Cambridge U.P. (2003).

\bibitem{D'Agostini}
G.~D'Agostini.
\newblock {"Bayesian reasoning in data analysis: A critical introduction"}.
\newblock World Scientific (2003).

\bibitem{Akaike}
Hirotugu Akaike.
\newblock {"A new look at the statistical model identification"}.
\newblock {\em IEEE Transactions on Automatic Control}, 16 (6):716--723, 1974.

\bibitem{Schwarz}
G.~Schwarz.
\newblock {"Estimating the dimension of a model"}.
\newblock {\em Annals of Statistics}, 6 (2):461--464, 1978.

\bibitem{Liddle2004}
Andrew~R. Liddle.
\newblock {"How many cosmological parameters?"}.
\newblock {\em Mon. Not. Roy. Astron. Soc.}, 351:L49--L53, 2004,

\bibitem{Beltran2005}
Maria Beltran, Juan Garcia-Bellido, Julien Lesgourgues, Andrew~R Liddle, and
  Anze Slosar.
\newblock {"Bayesian model selection and isocurvature perturbations"}.
\newblock {\em Phys. Rev.}, D71:063532, 2005, astro-ph/0501477.

\bibitem{Parkinson2006}
David Parkinson, Pia Mukherjee, and Andrew~R Liddle.
\newblock {"A Bayesian model selection analysis of WMAP3"}.
\newblock {\em Phys. Rev.}, D73:123523, 2006, astro-ph/0605003.

\bibitem{Trotta2007}
Roberto Trotta.
\newblock {"Applications of Bayesian model selection to cosmological
  parameters"}.
\newblock {\em Mon. Not. Roy. Astron. Soc.}, 378:72--82, 2007,
  astro-ph/0504022.

\bibitem{Kofman:1986wm}
L.~A. Kofman and Andrei~D. Linde.
\newblock {Generation of Density Perturbations in the Inflationary Cosmology}.
\newblock {\em Nucl. Phys.}, B282:555, 1987.

\bibitem{Chen:2008wn}
Xingang Chen, Richard Easther, and Eugene~A. Lim.
\newblock {Generation and Characterization of Large Non-Gaussianities in Single
  Field Inflation}.
\newblock 2008, arXiv:0801.3295 [astro-ph].

\bibitem{Piran}
T.~Paczynski, B.~and~Piran.
\newblock A dipole moment of the microwave background as a cosmological effect.
\newblock {\em Astrophys. J.}, 364:341--348, December 1990.

\bibitem{Alnes:2006pf}
Havard Alnes and Morad Amarzguioui.
\newblock {"CMB anisotropies seen by an off-center observer in a spherically
  symmetric inhomogeneous universe"}.
\newblock {\em Phys. Rev.}, D74:103520, 2006, astro-ph/0607334.

\bibitem{Rasanen:2003}
Syksy Rasanen.
\newblock {Dark energy from backreaction}.
\newblock {\em JCAP}, 0402:003, 2004, astro-ph/0311257.

\bibitem{Mattsson:2007tj}
Teppo Mattsson.
\newblock {"Dark energy as a mirage"}.
\newblock 2007, arXiv:0711.4264 [astro-ph].

\bibitem{Wiltshire:2007zj}
David~L. Wiltshire.
\newblock {Dark energy without dark energy}.
\newblock 2007, arXiv:0712.3984 [astro-ph].

\bibitem{Li:2008yj}
Nan Li, Marina Seikel, and Dominik~J. Schwarz.
\newblock {"Is dark energy an effect of averaging?"}.
\newblock 2008, arXiv:0801.3420 [astro-ph].

\end{thebibliography}

\end{document}